\documentclass[english,11pt,final,3p]{article}
\usepackage[T1]{fontenc}
\usepackage[latin9]{inputenc}
\usepackage{geometry}
\usepackage{xcolor}
\usepackage{array}
\usepackage{float}
\usepackage{booktabs}
\usepackage{mathrsfs}
\usepackage{multirow}
\usepackage{amsmath}
\usepackage{amssymb}
\usepackage{graphicx}
\usepackage[authoryear]{natbib}
\PassOptionsToPackage{normalem}{ulem}
\usepackage{ulem}

\makeatletter

\providecommand{\tabularnewline}{\\}
\providecolor{lyxadded}{rgb}{0,0,1}
\providecolor{lyxdeleted}{rgb}{1,0,0}

\DeclareRobustCommand{\lyxsout}[1]{\ifx\\#1\else\sout{#1}\fi}

\usepackage{graphicx}
\usepackage{subfigure}
\usepackage{amsmath, amsthm, amssymb}

\makeatother

\usepackage{babel}
\begin{document}
\title{Propagation of Elongated Fluid-Driven Fractures: \\
Rock Toughness vs. Fluid Viscosity\thanks{submitted for consideration for publication in ASME Journal of Applied Mechanics}}
\author{Dmitry I. Garagash}
\maketitle
\begin{center}
Dalhousie University, Department of Civil and Resource Engineering,
Halifax, Canada\\
garagash@dal.ca
\par\end{center}
\begin{abstract}
This paper studies the effect of the rock fracture toughness on the
propagation of elongated fluid-driven fractures. We use the `tough
PKN' model of \citet{SarvaraminiGaragash15}, an extension of the
classical PKN model \citep{PeKe61,Nord72}, which allows for a non-zero
energy release rate into the advancing fracture front(s). We provide
a self-consistent analysis of a `tough' elongated fracture driven
by arbitrary fluid injection law under assumption of the negligible
fluid leak-off. We use scaling considerations to identify the non-dimensional
parameters governing the propagation regimes and their succession
in time, provide a number of analytical solutions in the limiting
regimes for an arbitrary power-law injection, and also posit a simplified,
equation-of-motion, approach to solve a general elongated fracture
propagation problem during the injection and shut-in periods. Finally,
we use the developed solutions for a tough elongated fracture to surmise
the relative importance of the viscous and toughness-related dissipation
on the fracture dynamics and broach the implications of the possible
toughness scale-dependence.
\end{abstract}

\subsubsection*{Key points:}
\begin{enumerate}
\item Model for propagation of elongated hydraulic fractures accounting
for solid toughness and fluid viscosity for varied fluid injection
scenarios.
\item Analytical solutions for end-member toughness- and viscosity-dominated
regimes for a power-law and exponential fluid injection in time.
\item Solution to the transient fracture growth in mixed (toughness-viscosity)
regime by a numerical and a simplified analytical Equation-of-Motion
approaches.
\end{enumerate}

\section{Introduction}

\begin{figure}[p]
\begin{centering}
(a)\includegraphics[scale=0.42]{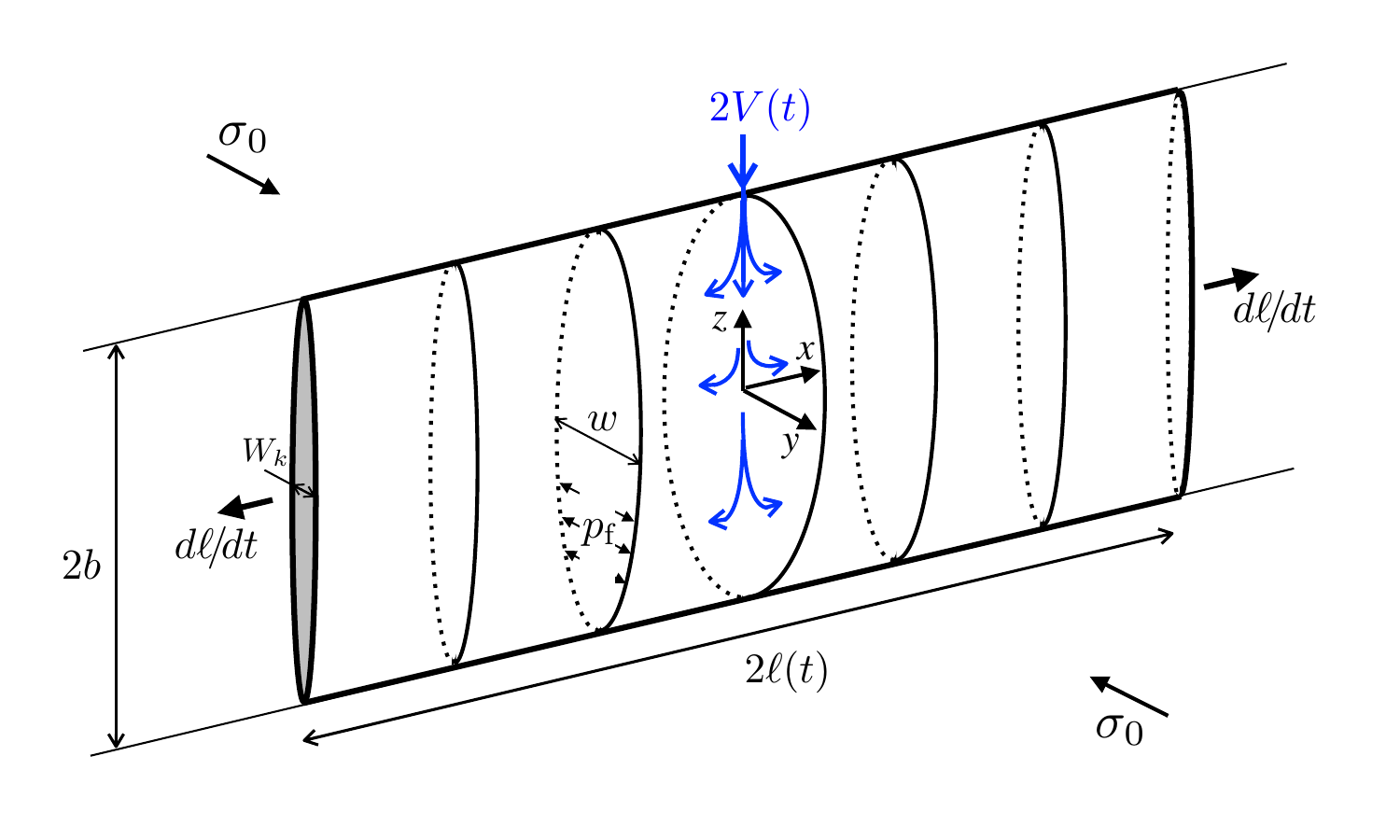}
\par\end{centering}
\begin{centering}
(b)\includegraphics[scale=0.32]{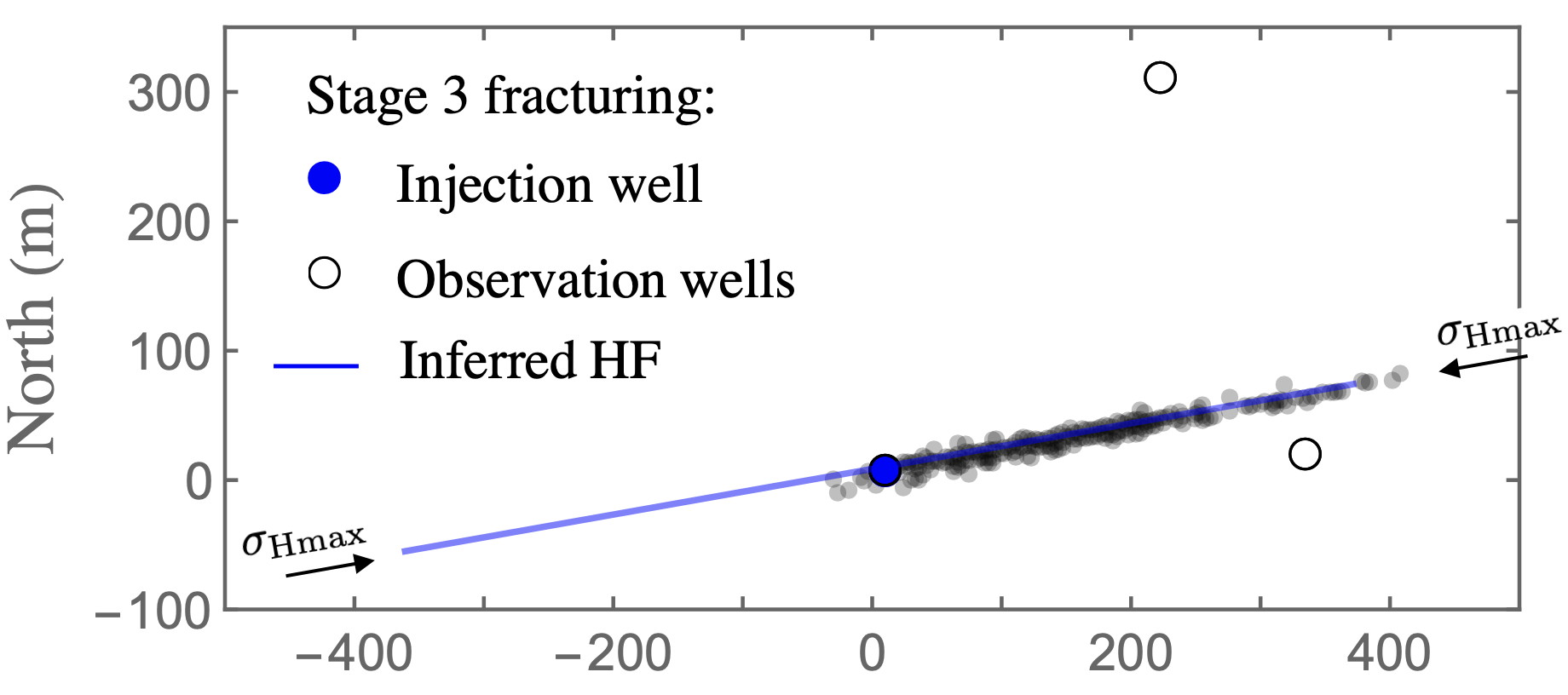}
\par\end{centering}
\begin{centering}
(c)\includegraphics[scale=0.16]{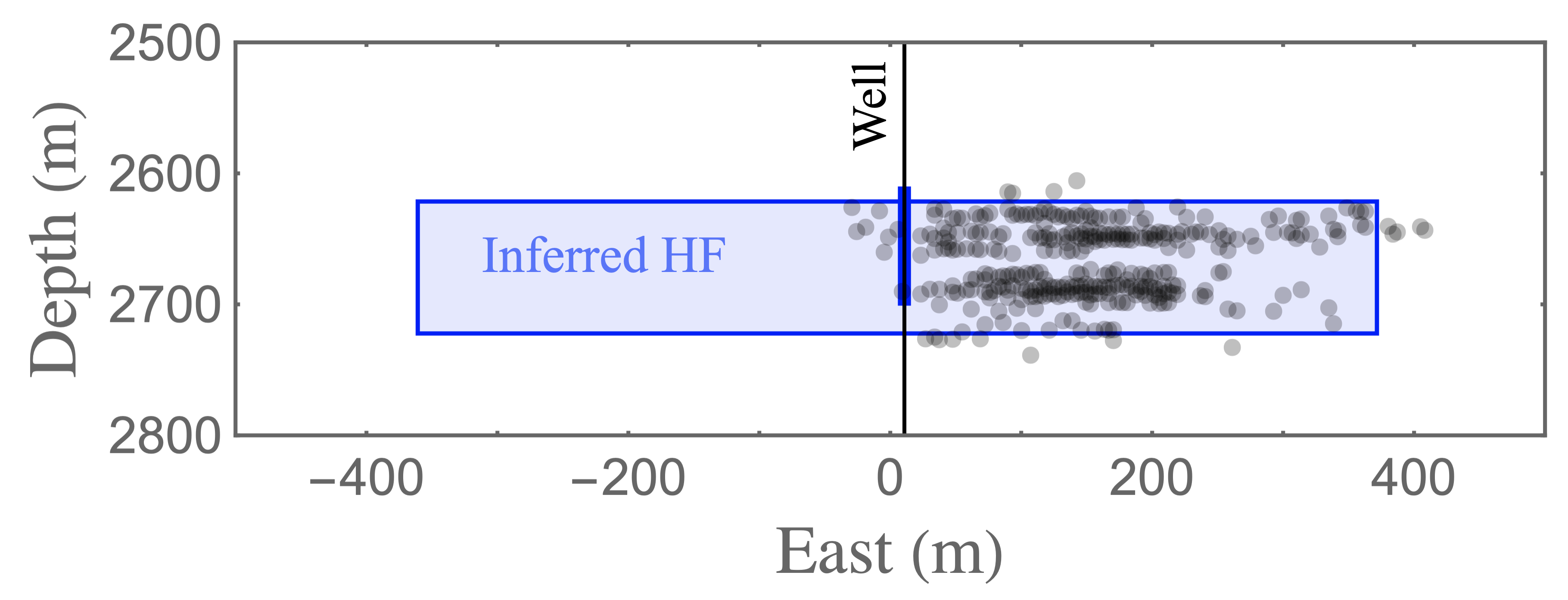}
\par\end{centering}
\caption{\textbf{(a)} Elongated, bi-wing hydraulic fracture of length $2\ell(t)$
`tunneling' in a layer of height $2b$, $b\ll\ell$, driven by fluid
injection with cumulative volume $2V(t)$. \textbf{(b-c)} Relocated
micro-seismicity in the top and side (EW) views during Stage 3 hydraulic
fracturing injection in vertical well 21-10 in Carthage Cotton Valley
Gas Field, Texas \citep[modified after][]{Rutledge04,Mayerhofer00}.
Microseismicity is shown by opaque gray dots, such that darker parts
of the micro-seismicity `cloud' correspond to higher spatial density
of events. Microseismicity, which is induced on natural fractures
along the path of the propagating hydraulic fracture, highlights the
east wing of an elongated hydraulic fracture with aspect ratio $b/\ell\approx1/8$,
(c), aligned in the direction N80E of the maximum regional horizontal
stress \citep{laubach1988coring}, (b). Lack of observed micro-seismicity
to the west of the injection well can be due to a distant (eastward)
location of the two observation wells in this study, thus, a symmetric
bi-wing fracture is assumed here. Rectangle in (c) shows the inferred
fracture footprint. Perforated well interval over which the fluid
injection took place is shown by a thick blue line in (c).\textbf{
} \label{fig1}}
\end{figure}

This paper studies propagation of three-dimensional hydraulic fractures
when one of its dimensions (e.g. height) remains constrained. These
`elongated' hydraulic fractures arise in geo-reservoirs stimulation
applications \citep{PeKe61,Nord72,AdSi06}, as well as, occur naturally,
such as in the joint formation in pore-fluid-overpressured sedimentary
strata \citep{PollardAydin88}, lateral magma emplacement in rift-zones
by blade-like dikes \citep{RubinPollard1987-BladeDikes,Rivalta15_diketoughness},
vertical magma transport in the crust by buoyant dikes with toughness
constrained breadth \citep{garagash2014gravity,garagash2022-gravityHF,Davis20_radialdike,MoriLecampion22}.
Fig. \ref{fig1}a shows a conceptualized bi-wing, elongated fracture
of height $2b$ and length $2\ell(t)$ with $\ell\gg b$ driven by
the volumetric fluid source $2V(t)$ at the center. Fig. \ref{fig1}b,c
show the relocated micro-seismicity recorded during propagation of
an industrial hydraulic fracture in Carthage Cotton Valley Gas Field,
Texas \citep{Rutledge04,Mayerhofer00}, highlighting the final spatial
extent of an elongated fracture. Lithological layering and related
stress barriers to hydraulic fracture vertical propagation are usually
evoked to explain the laterally-elongated (constrained height) fracture
in hydraulic fracturing in geo-energy applications \citep[e.g.,][]{adachi2010pseudo3D,Mori24_HF_StressBarrier}
and in formation of natural joints by pore fluid overpressure \citep{PollardAydin88},
while density-stratification of the crust in rifting zones and associated
reversal of the magma buoyancy in shallow crust can lead to the formation
and sustained lateral growth of blade-like dikes \citep{RubinPollard1987-BladeDikes,Lister90b,townsend2017-BladeDike}.

\subsection{Elongated hydraulic fracture modeling}

The classical PKN \citep{PeKe61,Nord72} model of elongated hydraulic
fracturing is based on \emph{two main assumptions}. 
\begin{enumerate}
\item The crack length is much larger than the height, which allows to neglect
elastic stress transfer along the elongated crack dimension and thus
model a vertical fracture cross-section as an uniformly-pressurized,
one-dimensional Griffith crack (Fig. \ref{fig1}a). 
\item The crack is `closed' at the advancing fracture edges $x=\pm\ell(t)$,
i.e. the crack opening is assumed to be zero there. Although seemingly
physically sensible, when taken together with the first, negligible-stress-transfer
assumption, the `closed-edge' condition can be shown to correspond
to the energy release rate into the advancing fracture front, thus
rendering the classical PKN - a \emph{zero-toughness} model of an
elongated fracture. 
\end{enumerate}
Early studies of the classical PKN model \citep{Nord72,Kemp90} have
been recently revisited and expanded upon by \citep{Kovalyshen10-PKN,mishuris2012modeling,WrobelMishuris15}.

While the zero-stress-transfer assumption is approximately valid in
the `body' of the elongated fracture, it does break down near the
advancing edges $x=\pm\ell(t)$ where the actual fracture geometry
is more adequately approximated by a transition from the plane-strain
(`infinite'-height), fully-elastically-coupled crack geometry at small
distances from an edge $|\ell-x|\ll b$ to the PKN crack geometry
away from the edges $|\ell-x|\gg b$ \citep{AdachiPeirce08,dontsov2016-PKN-pseudo3D,peruzzo2024contained}.
Fortunately, the near-front complications of the full solution are
not important on the scale of the entire fracture, as long as the
properly reduced front propagation condition(s) for an elongated fracture
are imposed. These conditions have been robustly established by \citet{SarvaraminiGaragash15},
who derived the energy release into the near-front region of an elongated
crack and provided the first unambiguous extension of the classic
PKN hydraulic fracture model to account for a non-zero fracture toughness.
This model is hereafter referred to as the `tough PKN' or `tough elongated'
fracture model. In an earlier ad hoc treatment of the toughness problem,
\citet{Nolte91} proposed to `fit' a half-penny-shape crack to the
front region of an elongated fracture and to constrain the fluid pressure
at the crack tip by requiring the stress intensity factor of the penny-shape
crack to match the rock toughness. \citet{RubinPollard1987-BladeDikes,townsend2017-BladeDike}
employed a similar ad hoc approach to model effect of toughness on
the lateral propagation of blade-like dikes. \citet{dontsov2016-PKN-pseudo3D,peruzzo2024contained}
validated the `tough PKN' model of \citet{SarvaraminiGaragash15}
by explicit numerical solutions for the elongated hydraulic fracture
propagation with spontaneously evolving front(s). Outside of the realm
of fluid-driven fractures, the energy-release-based approach to propagation
of elongated cracks have been used to model `tunneling' fractures
in layered material composites \citep[and references therein]{HuSu92}
and large earthquake (shear) ruptures \citep{weng2019dynamics}. 

\citet{chuprakov2017-PKN} has used the tough PKN model of \citet{SarvaraminiGaragash15}
to investigate numerically the hydraulic fracture propagation after
the injection shut-in, when the rock toughness and the fluid leak-off
effectively govern the slowing-down fracture propagation towards the
ultimate arrest. \citet{dontsov22PKN} have provided a complete analysis
of the propagation of a tough elongated hydraulic fracture in the
case of a constant rate of fluid injection and accounting for the
fluid leak-off, further extending this work to the case of non-Newtonian
fracturing fluid in \citep{dontsov2022PKN_non-Newtonian}. Here, we
extend these previous studies to provide a self-consistent analysis
of a `tough' elongated fracture driven by arbitrary injection law
under assumption of negligible fluid leak-off. We use scaling considerations
to identify the non-dimensional parameters governing the propagation
regimes and their succession in time, provide a number of analytical
solutions in the limiting regimes for arbitrary power-law injection,
and posit a simplified, equation-of-motion, approach to solve a general
elongated fracture propagation problem. 

\subsection{Rock fracture toughness and scale dependence}

 It has been long suggested that fracture toughness $K_{Ic}$ (or
corresponding fracture energy $G_{c}=K_{Ic}^{2}/E'$ where $E'$ is
the the plane-strain modulus) scales with fracture size. A number
of observations, including that of laboratory fracture (see \citet{LiuLecampionGaragash19}
for review), field observations of damage zones abating magmatic dikes
\citep{delaney1986DikeJoints} and reservoir hydraulic fractures \citep{warpinski1993examination},
and field inferences based on the scaling of the maximum crack opening
with size for veins and dikes \citep[and references therein]{Scholz10,Rivalta15_diketoughness}
and reservoir hydraulic fractures \citep{shlyapobersky1985energy,warpinski1998GRI_FieldExperiment_Toughness}
can be formally encapsulated in a power law relation 
\begin{equation}
K_{Ic}=K_{Ic,0}\left(\frac{b}{b_{0}}\right)^{\chi/2},\quad\text{or equivivalently}\quad G_{c}=G_{c,0}\left(\frac{b}{b_{0}}\right)^{\chi}\label{tough}
\end{equation}
where $K_{Ic,0}$ and $G_{c,0}=K_{Ic,0}^{2}/E'$ are the reference
toughness and fracture energy values at a reference fracture size
$b_{0}$, and $\chi\ge0$ is the scaling exponent. A particular measure
of fracture size $b$ in (\ref{tough}) is taken as the fracture intermediate
dimension for an elongated fracture, i.e. half-height for a laterally-propagating
(Fig. \ref{fig1}) or half-breadth for a vertically propagating fracture
(e.g. buoyant dikes), respectively. This measure of the fracture size
in scaling relation (\ref{tough}) can be generalized to an arbitrary
planar fracture geometry, for example, by taking $b$ as the (local)
radius of curvature of the propagating fracture edge. 

Experiments constrain toughness of clastic rocks $K_{Ic,0}\sim0.5$
to $1$ $\text{MPa\,m}^{1/2}$ and $G_{c,0}\sim10$ to $30$ $\text{Pa m}$
on the laboratory scale $2b_{0}\sim0.1$ m \citep[and reference therein]{schmidt1977toughness,nara2012toughness,chandler2016fracture,noel2021toughness}.
Values of the power-law scaling exponent inferred for geomaterials
and wood over a limited range of fracture sizes in a laboratory are
$\chi\approx0.4$ to $0.6$ \citep{morel2002r,lopez1998anomalous}),
while $\chi\approx1$ has been inferred on a much larger scale in
the field \citep[e.g., ][]{Scholz10}. Such parametrized relation
(\ref{tough}) is consistent with the \emph{one-to-two order of magnitude
increase} in toughness from the lab to the field scale inferred for
geo-reservoir hydraulic fracturing \citep{shlyapobersky1985energy,warpinski1998GRI_FieldExperiment_Toughness}
and magmatic dikes \citep[e.g., ][]{Scholz10,Rivalta15_diketoughness}.
It must also be said, however, that the field inferences of the rock
fracture toughness may actually lump together the effects of the rock
toughness and of the dissipation in the viscous fluid flow in the
fracture, thus overestimating the former \citep[e.g.,][]{Rivalta15_diketoughness,liu2022HFtough}. 

In this study, we develop solutions to a physically-sound model of
a `tough' elongated hydraulic fracture that is not limited to a single
dissipation mechanism, and thus, when combined with observations,
may shed further light on the fracture toughness scaling and its implications
to industrial and natural fracturing.

\subsection{Paper organization}

The paper is organized as follows. Section 2 introduces governing
equations for a fluid-driven crack tunneling in a layer of fixed height.
Section 3 discusses the asymptotics of the solution near the fracture
front, emergent limiting fracture propagation regimes, and corresponding
regime motivated scaling of the problem. Section 4 presents self-similar
solutions for the fracture propagation in the limiting toughness-
and viscosity- dominated regimes of propagation for a power-law and
exponential fluid injection (including the cases of constant injection
volumetric rate, constant injection pressure, and constant injected
volume). Section 5 addresses a general transient propagation problem
with arbitrary injection schedule by using a simplified equation-of-motion
solution method, as well as, a full numerical solution approach. Solution
for the shut-in problem is furnished. In Section 6, we discuss the
developed solutions to surmise the relative importance of the viscous
and toughness related dissipation on the dynamics of an elongated
hydraulic fracture during the injection and shut-in stages, respectively,
and broach the implications of the toughness scale dependence (\ref{tough})
on the fluid-driven fracture dynamics.  

\section{Governing Equations for Tough Elongated Fluid-Driven Crack}

Consider propagation of a hydraulic fracture driven by the injection
of fluid volume $2V(t)$ and constrained in height to a layer of thickness
$2b$ (Fig. \ref{fig1}a). The crack is internally loaded by distributed
fluid pressure $p_{f}(x)$ and remotely confined by minimum in-situ
compressive stress $\sigma_{o}$. The fracture geometry is therefore
set by the expanding crack length $2\ell(t)$ and constant height
$2b$. For an elongated crack, $b\ll\ell$, crack opening in a given
$x$ cross-section is approximated by the Griffith's solution for
a plane-strain, uniformly-pressurized crack, $w(x,z)=(4p(x)/E^{\prime})\sqrt{b^{2}-z^{2}}$,
where $E'=E/(1-\nu^{2})$ is the plane-strain elastic modulus. Unidirectional
flow inside the crack of an incompressible fluid with viscosity $\mu$
is governed by Poiseuille\textquoteright s law, $v(x,z)=-(w^{2}(x,z)/12\mu)(\partial p/\partial x)$,
and continuity,
\begin{equation}
\frac{\partial w}{\partial t}+\frac{\partial wv}{\partial x}=0.\label{cont}
\end{equation}

We define height-averages $w(x)$ and $v(x)$ of the local crack opening
$w(x,z)$ and fluid velocity $v(x,z)$
\begin{equation}
w(x)=\frac{1}{2b}\int w(x,z)dz=\frac{bp(x)}{\bar{E}},\qquad v(x)=\frac{1}{2bw(x)}\int w(x,z)v(x,z)dz=-\frac{w^{2}(x)}{\bar{\mu}}\frac{\partial p}{\partial x}\label{W}
\end{equation}
in terms effective modulus $\bar{E}=E'/\pi$ and viscosity $\bar{\mu}=\pi^{2}\mu$.
 Substituting expression for $p$ from the first into the second
equation in the above, we get a convenient form of the lubrication
equation for the tunneling crack
\begin{equation}
v(x)=-\frac{\bar{E}}{3\bar{\mu}b}\frac{\partial w^{3}(x)}{\partial x}\label{lub}
\end{equation}

Boundary conditions at the fracture tip $x=\ell(t)$ and inlet $x=0$
for the above set of field equations are formulated in the following,
while making use of the problem symmetry about the inlet.
\begin{itemize}
\item Fluid continuity at the tip, 
\begin{equation}
x=\ell:\quad v=\dot{\ell}\label{tip}
\end{equation}
Here $\dot{\ell}=d\ell/dt$ is the propagation velocity.
\item Fluid continuity at the inlet,
\begin{equation}
x=0:\quad2bwv=\dot{V}\label{inlet}
\end{equation}
Here $V$ and $\dot{V}=dV/dt$ are the (half) injection volume and
its rate, respectively. Inlet condition (\ref{inlet}) can be replaced
by the global continuity statement
\begin{equation}
\int_{0}^{\ell}2bwdx=V\label{global}
\end{equation}
\item \citet{SarvaraminiGaragash15} established that elastic energy release
$dU$ to advance the tunneling fracture front by $d\ell>0$ is approximately
$dU\approx U_{2D}(\ell)d\ell$, where  $U_{\text{2D}}(\ell)=b^{2}p^{2}(\ell)/\bar{E}$
is the internal energy of a two-dimensional (plain-strain, Griffith)
crack of extent $b$ uniformly pressurized by net-pressure value $p(\ell)$
at the front of the tunneling crack. The front-average elastic energy
release rate is then $G=dU/(2bd\ell)=bp^{2}(\ell)/(2\bar{E})$. Quasi-static
fracture propagation requires that the energy release rate $G$ is
equal to the fracture energy $G_{c}$, which can be related to the
more-commonly used solid toughness $K_{Ic}$ by $G_{c}=K_{Ic}^{2}/E^{\prime}$.
In view of the above expression for $G$, the propagation condition
$G=G_{c}$ constrains the value of the net-pressure at the fracture
front to the critical value $P_{k}$,
\begin{equation}
x=\ell:\quad p=P_{k}\equiv\frac{\bar{K}}{\sqrt{b}},\label{prop_P}
\end{equation}
where $\bar{K}=\sqrt{2/\pi}K_{Ic}$ is an effective toughness parameter.
Alternatively, the above propagation condition can be expressed in
terms of the opening, see the first in (\ref{W}), attaining a critical
value $W_{k}$ at the fracture front:
\begin{equation}
x=\ell:\quad w=W_{k}\equiv\frac{\bar{K}\sqrt{b}}{\bar{E}}.\label{prop}
\end{equation}
 The opening at the front of an elongated fracture is therefore finite
as long as the material toughness is. This apparently non-physical
condition is reconciled by considering that the presented model is
an adequate approximation for an elongated fracture starting only
from some distance away from the propagating edge, more specifically
at distances greater than the crack height $2b$. In other words,
the model provides an `outer' solution to the elongated crack problem,
i.e. the solution outside of the immediate vicinity of the propagating
tips, which does not have to satisfy the physical constraints at
the actual fracture tip, e.g. that of zero opening there. Full numerical
solutions of hydraulic fracture propagation with constrained heigh
\citep{dontsov2016-PKN-pseudo3D,peruzzo2024contained}\footnote{where fracture height growth is not precluded, but limited by the
presence of the stress barriers, such that the blade-like geometry
$b(x,t)\ll\ell(t)$ is maintained.} validate the applicability of the simplified elongated crack model
with propagation condition (\ref{prop_P}). We also observe that the
PKN model (which is equivalent to equations (\ref{lub}-\ref{global})
with the additional condition of zero crack opening at the propagating
edges), is recovered from the current model in the limit of zero toughness.
\end{itemize}
In summary, propagation of an elongated hydraulic fracture is governed
by continuity (\ref{cont}) and lubrication (\ref{lub}) phrased in
terms of height-averaged $w$ and $v$, (\ref{W}), and the boundary
conditions at the fracture tip, (\ref{tip}) and (\ref{prop}), and
the inlet, (\ref{inlet}) or (\ref{global}). Solution of the model
is parametrized by fracture height $2b$, injected fluid volume history
$2V(t)$ and the three effective material parameters (elastic modulus,
fluid viscosity, and toughness), 
\begin{equation}
\bar{E}=\frac{1}{\pi}\frac{E}{1-\nu^{2}},\quad\bar{\mu}=\pi^{2}\mu,\quad\bar{K}=\sqrt{\frac{2}{\pi}}K_{Ic}.\label{params}
\end{equation}

\section{Tip Asymptotics and Scaling}

Integrating lubrication (\ref{lub}) near the fracture tip, where
$v\approx\dot{\ell}$, (\ref{tip}), and further applying propagation
condition, (\ref{prop}), allows to express the crack opening asymptote
there as,
\begin{equation}
x\rightarrow\ell:\quad w=\left(W_{k}^{3}+W_{m}^{3}\frac{\ell-x}{\ell}\right)^{1/3},\label{asy}
\end{equation}
where $W_{k}$, given in (\ref{prop}), and 
\begin{equation}
W_{m}=\left(\frac{3\bar{\mu}b}{\bar{E}}\,\ell\,\dot{\ell}\right)^{1/3}\label{Wm}
\end{equation}
can be identified as characteristic values of the crack opening under
the toughness and viscosity dominated conditions, respectively. Indeed,
setting $\bar{\mu}=0$ in the \textit{toughness-dominated regime},
we have $w(x,t)=W_{k}$ near the tip (and, as shown further, everywhere
along the crack), while setting $K_{Ic}=0$ in the \textit{viscosity-dominated
regime}, we recover $w(x,t)=W_{m}(t)\,\left(1-x/\ell(t)\right)^{1/3}$
near the tip \citep{Kemp90,Kovalyshen10-PKN}. Therefore, in general,
these two limiting regimes can be identified by comparing magnitudes
of $W_{k}$ and $W_{m}$. It follows from definitions (\ref{prop})
and (\ref{Wm}), that, when neither fluid viscosity nor the solid
fracture toughness are exactly zero and since $W_{k}$ is independent
of time, fracture evolves towards the viscosity (toughness) dominated
regime if the crack grows faster (slower) than $\ell\sim\sqrt{t}$,
i.e., $W_{m}$ is increasing (decreasing) function of time. General
problem of the crack evolution can be therefore cast in terms of the
transition between these two limiting regimes.

It is useful for further solution development to introduce two scaling
schemes for the general solution, as suggested by the toughness- and
viscosity- dominated end-member cases introduced in the above. Let
us define normalized opening/net-pressure $\Omega$, fluid velocity
$\vartheta$, and crack length $\gamma$ as functions of normalized
coordinate $\xi=x/\ell(t)$ and time: 
\begin{equation}
w(x,t)=(b/\bar{E})\,p(x,t)=W(t)\,\Omega(\xi,t),\quad v(x,t)=\dot{\ell}(t)\,\vartheta(\xi,t),\quad\ell(t)=L(t)\gamma(t)\label{scaling}
\end{equation}
with the characteristic opening $W$ and length $L$ scales given
by 
\begin{equation}
W(t)=\left\{ \begin{array}{ll}
W_{k}=\bar{K}\sqrt{b}/\bar{E} & \text{(toughness scaling)}\\
W_{m}(t)=\left(3\bar{\mu}b\,\ell\,\dot{\ell}/\bar{E}\right)^{1/3}\  & \text{(viscosity scaling)}
\end{array}\right.,\quad L(t)=\frac{V(t)}{2b\,W(t)},\label{L}
\end{equation}
We note that since $W_{m}$ is defined in terms of the unknown crack
length $\ell$, the corresponding expression (\ref{L}) for the characteristic
length $L_{m}$ in the viscosity scaling is implicit, and furthermore
is dependent upon the complete solution. 

Substituting (\ref{scaling})-(\ref{L}) into the local continuity
(\ref{cont}) yields the normalized equation
\begin{equation}
t\dot{\Omega}+\frac{t\dot{W}}{W}\Omega-\frac{t\dot{\ell}}{\ell}\left(\xi\frac{\partial\Omega}{\partial\xi}-\frac{\partial\Omega\vartheta}{\partial\xi}\right)=0\label{cont'}
\end{equation}
or, in alternative form, obtained by integrating in space, using crack
tip continuity, $\vartheta{}_{\left|\xi\rightarrow1\right.}=1$, and
resolving for the fluid velocity,
\begin{equation}
\vartheta=\xi-\frac{1}{\Omega}\int_{1}^{\xi}\left(\Omega+\frac{t\dot{\Omega}+\frac{t\dot{W}}{W}\Omega}{t\dot{\ell}/\ell}\right)d\xi\label{cont''}
\end{equation}
where
\begin{equation}
\frac{t\dot{\ell}}{\ell}=\frac{t\dot{L}}{L}+\frac{t\dot{\gamma}}{\gamma}=\frac{t\dot{V}}{V}-\frac{t\dot{W}}{W}+\frac{t\dot{\gamma}}{\gamma}\label{L'}
\end{equation}
and the time derivatives in the normalized equations are carried out
at a fixed $\xi$. 

Normalized form of the fluid continuity at the inlet (\ref{inlet})
is 
\begin{equation}
\left(\Omega\vartheta\right)_{\left|\xi=0\right.}=\frac{t\dot{V}/V}{t\dot{\ell}/\ell}\frac{1}{\gamma}\label{inlet'}
\end{equation}
while, equivalently, the normalized form of global continuity (\ref{cont})
is
\begin{equation}
\int_{0}^{1}\Omega d\xi=\frac{1}{\gamma}\label{global'}
\end{equation}

Normalized lubrication equation (\ref{lub}) and the tip asymptote
(\ref{asy}), which embodies both the tip fluid continuity (\ref{tip})
and the tip propagation (\ref{prop}) conditions, in the two scalings
read: 
\begin{equation}
\text{toughness scaling:}\quad\vartheta=-\frac{1}{\mathbb{M}}\frac{\partial\Omega^{3}}{\partial\xi},\quad\Omega_{\left|\xi\rightarrow1\right.}=\left(1+\mathbb{M}\,(1-\xi)\right)^{1/3},\label{Ktip}
\end{equation}
\begin{equation}
\text{viscosity scaling:}\quad\vartheta=-\frac{\partial\Omega^{3}}{\partial\xi},\quad\Omega_{\left|\xi\rightarrow1\right.}=\left(\mathbb{K}^{3}+1-\xi\right)^{1/3}.\label{Mtip}
\end{equation}
where normalized crack front parameters $\mathbb{K}$ and $\mathbb{M}$
are defined in terms of the ratio of the two characteristic opening
scales (\ref{L}),
\begin{equation}
\mathbb{K}=\frac{W_{k}}{W_{m}}=\frac{\bar{K}\,b^{1/6}}{\left(3\bar{\mu}\bar{E}^{2}\,\ell\,\dot{\ell}\right)^{1/3}},\qquad\mathbb{M}=\frac{W_{m}^{3}}{W_{k}^{3}}=\frac{3\bar{\mu}\bar{E}^{2}\,\ell\,\dot{\ell}}{\bar{K}^{3}b^{1/2}}.\label{MK}
\end{equation}

Non-dimensional crack opening, half-length, and fluid velocity in
the two scalings are simply inter-related, 
\begin{equation}
\frac{\Omega_{m}}{\Omega_{k}}=\frac{\gamma_{k}}{\gamma_{m}}=\frac{W_{k}}{W_{m}}=\frac{L_{m}}{L_{k}}=\mathbb{K}=\mathbb{M}^{-1/3},\quad\frac{\vartheta_{m}}{\vartheta_{k}}=1\label{conv}
\end{equation}
where indices refer to a particular scaling used. 

Nondimensional parameters (\ref{MK}) characterize the relative importance
of the corresponding energy dissipation mechanisms in hydraulic fracturing
with constrained height, such that one can formally define the viscosity-dominated
($\mathbb{K}\ll1$) and the toughness-dominated ($\mathbb{M}\ll1$)
propagation regimes. 

\subsection{Alternate, Explicit Viscosity Scaling}

Viscosity scaling introduced in the above is implicit, since the scales
for the opening, $W_{m}$, and length, $L_{m}$, depend on the unknown
crack half-length and its time derivative (propagation speed). It
will therefore be useful to introduce an alternative, explicit set
of scales, $\overline{W}_{m}$ and $\overline{L}_{m}$, which are
independent of the sought solution, by writing
\begin{equation}
W_{m}=(t\dot{\ell}/\ell)^{1/5}\gamma_{m}^{2/5}\:\overline{W}_{m},\qquad L_{m}=(t\dot{\ell}/\ell)^{-1/5}\gamma_{m}^{-2/5}\,\overline{L}_{m}\label{W_m}
\end{equation}
where
\begin{equation}
\overline{W}_{m}=\left(\frac{3}{4}\frac{\bar{\mu}V^{2}(t)}{\bar{E}b\,t}\right)^{1/5},\qquad\overline{L}_{m}=\frac{V(t)}{2b\overline{W}_{m}}=\left(\frac{\bar{E}V^{3}(t)\,t}{24\bar{\mu}b^{4}}\right)^{1/5},\label{W_m_bar}
\end{equation}

Using (\ref{W_m}) in (\ref{MK}), we can write for the implicit normalized
toughness $\mathscr{\mathbb{K}}$ and viscosity $\mathbb{M}$ parameters
\begin{equation}
\mathscr{\mathbb{K}}=(t\dot{\ell}/\ell)^{-1/5}\,\gamma_{m}^{-2/5}\,\mathscr{\mathcal{K}},\qquad\mathcal{\mathbb{M}}=(t\dot{\ell}/\ell)\,\gamma_{k}^{2}\,\mathcal{M},\label{K}
\end{equation}
in terms of alternative explicit normalized toughness $\mathscr{\mathcal{K}}$
and viscosity $\mathcal{M}=\mathscr{\mathcal{K}}^{-5}$ parameters
defined as
\begin{equation}
\mathscr{\mathcal{K}}=\frac{W_{k}}{\overline{W}_{m}}=\left(\frac{4}{3}\frac{\bar{K}^{5}b^{7/2}\,t}{\bar{\mu}\bar{E}^{4}V^{2}(t)}\right)^{1/5},\qquad\mathcal{M}=\left(\frac{\overline{W}_{m}}{W_{k}}\right)^{5}=\frac{3}{4}\frac{\bar{\mu}\bar{E}^{4}V^{2}(t)}{\bar{K}^{5}b^{7/2}\,t}.\label{K_exp}
\end{equation}

In (\ref{W_m}) and (\ref{W_m}), $\gamma_{m,k}=\ell/L_{m,k}$ is
the previously-defined, normalized half-length in either the implicit-viscosity
($m$) or toughness ($k$) scaling.

Using explicit viscosity scales to define the corresponding normalized
opening/net pressure and crack half-length similar to (\ref{scaling}),
\begin{equation}
w(x,t)=(b/\bar{E})\,p(x,t)=\overline{W}_{m}(t)\,\bar{\Omega}_{m}(\xi,t),\quad\ell(t)=\overline{L}_{m}(t)\bar{\gamma}_{m}(t),\label{scaling-m-bar}
\end{equation}
the normalized solutions in the implicit and explicit viscosity scalings
can then be related as
\begin{equation}
\frac{\gamma_{m}}{\bar{\gamma}_{m}}=\frac{\bar{\Omega}_{m}}{\Omega_{m}}=\frac{\mathscr{\mathcal{K}}}{\mathscr{\mathbb{K}}}=(t\dot{\ell}/\ell)^{1/3}\,\bar{\gamma}_{m}^{2/3}\label{relation_m}
\end{equation}

\section{Limiting Self-Similar Solutions\label{Sec:regimes}}

\subsection{Zero-Viscosity Solution\label{Sec:inf}}

In the toughness-dominated regime, the solution is given by the \emph{zero-viscosity
solution} of the set of equations (\ref{cont}-\ref{prop}) or, in
the normalized form, (\ref{cont'})-(\ref{Ktip}). The solution is
particularly simple,- it is given in the \emph{toughness} \emph{scaling}
by 
\[
\mathbb{M}=0:\quad\Omega_{k}(\xi)=\gamma_{k}=1.
\]
where subscript \emph{k }points to the scaling used. In the dimensional
form,
\begin{equation}
\mu=0:\quad w(x,t)=(b/\bar{E})\,p(x,t)=W_{k},\quad\ell(t)=L_{k}(t)=\frac{V(t)}{2bW_{k}},\label{M0}
\end{equation}
it corresponds to uniformly pressurized and open crack, which extent
is proportional to the injected fluid volume.

It also proves convenient to formulate the \emph{small viscosity correction}
to the above zero-viscosity solution by considering the first term
in the solution expansion in $\mathbb{M}\ll1$. Using standard asymptotic
methods (see, e.g., a similar treatment of the plane-strain hydraulic
fracture \citep{Garagash06b}), we can find for power-law fluid injection,
$V\propto t^{\alpha}$,
\begin{equation}
\text{injection, }\alpha>0:\quad\Omega_{k}\approx\left(1+\mathbb{M}(1-\xi)\right)^{1/3},\quad\gamma_{k}\approx1-\frac{1}{6}\mathbb{M},\quad\mathbb{M}\approx\alpha\mathcal{M}\label{M1_inj}
\end{equation}
\begin{equation}
\text{fixed-volume, }\alpha=0:\quad\quad\Omega_{k}\approx\left(1+\mathbb{M}\frac{2}{\pi}\cos\frac{\pi\xi}{2}\right)^{1/3},\quad\gamma_{k}\approx1-\frac{4}{3\pi^{2}}\mathbb{M},\quad\mathbb{M}\sim\exp\left(-\frac{3\pi^{2}}{4\mathcal{M}}\right)\label{M1_shut}
\end{equation}

\subsection{Zero-Toughness Solution\label{Sec:zero}}

\subsubsection*{Similarity Considerations in Viscosity Scaling\label{Sec:similarity}}

Normalized equations in \emph{viscosity scaling}, (\ref{cont'})-(\ref{global'})
and (\ref{Mtip}), admit self-similar solutions, i.e. $\Omega_{m}=\Omega_{m}(\xi)$
and $\gamma_{m}=\text{{const}}$, when \textbf{(}\textbf{\emph{i}}\textbf{)}
dimensionless toughness parameter $\mathbb{K}$ is time-independent;
and \textbf{(}\textbf{\emph{ii}}\textbf{)} fluid injection is either
a power-law or an exponential of time.

Focusing on a \emph{power-law injection} first, the explicit scales
(\ref{W_m_bar}) and toughness parameter (\ref{K_exp}) are also power-laws,
\begin{equation}
V\propto t^{\alpha}:\quad\overline{L}_{m}\propto t^{(3\alpha+1)/5},\quad\overline{W}_{m}\propto t^{(2\alpha-1)/5},\quad\mathcal{K}\propto t^{-(2\alpha-1)/5},\label{power}
\end{equation}
and, in view of self-similarity ($\gamma_{m}=\text{{const}}$), so
are the corresponding implicit scales (\ref{W_m}) and toughness parameter
(\ref{K})
\begin{equation}
V\propto t^{\alpha}:\quad\frac{L_{m}}{\overline{L}_{m}}=\frac{\overline{W}_{m}}{W_{m}}=\frac{\mathbb{K}}{\mathcal{K}}=\left(\frac{3\alpha+1}{5}\right)^{-1/5}\gamma_{m}^{-2/5}=\text{const}.\label{power'}
\end{equation}
(Here we have used the similarity notion $\dot{\gamma}_{m}=0$ to
reduce (\ref{L'}) to $t\dot{\ell}/\ell=t\dot{L}_{m}/L_{m}=t\dot{\overline{L}}_{m}/\overline{L}_{m}=(3\alpha+1)/5$).
These allow to reduce the continuity, (\ref{cont'}), and lubrication,
the first in (\ref{Mtip}), equations to an ordinary differential
equation (ODE) in $\xi$,
\begin{equation}
\frac{2\alpha-1}{3\alpha+1}\Omega_{m}-\left(\xi\frac{d\Omega_{m}}{d\xi}-\frac{d\Omega_{m}\vartheta_{m}}{d\xi}\right)=0\quad\text{with}\quad\vartheta_{m}=-\frac{d\Omega_{m}^{3}}{d\xi},\label{lub''}
\end{equation}
This equation can be solved together with boundary conditions, $\Omega_{m}(1)=\mathbb{K}$
and $\vartheta_{m}(1)=1$, for the normalized opening distribution
$\Omega_{m}(\xi)$. The normalized half-length can then be obtained
by integrating (\ref{global'}), $\gamma_{m}=1/\int_{0}^{1}\Omega_{m}(\xi)d\xi$. 

Requirement of time-invariance of the dimensionless toughness $\mathbb{K}$,
i.e. self-similarity condition (\emph{i}) in the above, is satisfied
when either 
\begin{itemize}
\item $\mathcal{\mathbb{K}}=0$, i.e. when material toughness is zero $K_{Ic}=0$
and injection law is arbitrary, or 
\item $\mathcal{\mathbb{K}}=\text{const}$, when material toughness $K_{Ic}$
is arbitrary and injection law is particular, $V(t)\propto t^{\alpha}$
with $\alpha=1/2$. This case corresponds to \emph{injection at constant
inlet fluid pressure}, $p(0)=$const, constant inlet opening, $w(0)=\text{const}$,
and square-root-time expansion of the crack front $\ell\propto t^{1/2}$
- see (\ref{power}) for the time-dependence of the corresponding
scales.
\end{itemize}
For an \emph{exponential injection}, we can establish for the implicit
scales $L_{m}$ and $W_{m}$, and the toughness parameter $\mathbb{K}$
relationships similar to those for the power-law case in the above,
\begin{equation}
V\propto e^{t/t_{*}}:\quad\frac{L_{m}}{\overline{L}_{m}^{*}}=\frac{\overline{W}_{m}^{*}}{W_{m}}=\frac{\mathbb{K}}{\mathcal{K}^{*}}=\left(\frac{3}{5}\right)^{-1/5}\gamma_{m}^{-2/5}=\text{const}.\label{exp}
\end{equation}
where the newly defined explicit time-dependent scales $\overline{L}_{m}^{*}$
and $\overline{W}_{m}^{*}$ and toughness $\mathcal{K}^{*}$ for the
exponential injection case are
\begin{equation}
\overline{L}_{m}^{*}(t)=\overline{L}_{m}(t_{*})\,e^{3(t/t_{*}-1)/5},\quad\overline{W}_{m}^{*}(t)=\overline{W}_{m}(t_{*})\,e^{2(t/t_{*}-1)/5},\quad\mathcal{K}^{*}(t)=\mathcal{K}(t_{*})\,e^{-2(t/t_{*}-1)/5}\label{scales_exp}
\end{equation}
Here $t_{*}>0$ is injection timescale and the self-similarity is
assumed, i.e. $\gamma_{m}$ is a time-independent constant. Above
scaling relations show that length and opening evolve with time as
$\sim e^{3t/5t_{*}}$ and $\sim e^{2t/5t_{*}}$, respectively, while
dependence on time of the dimensionless toughness is reversed $\sim e^{-2t/5t_{*}}$.
Self-similarity, which requires $\mathbb{K}$ to be a time-invariant
constant, can therefore only be achieved when that constant is zero,
$\mathbb{K}=0$. This is realized for all times when material toughness
is zero $K'=0$ or asymptotically at times $t\gtrsim t_{*}$ when
$K_{Ic}>0$. In view of (\ref{exp}), local continuity and lubrication
can be shown to be formally equivalent to the power-law ones (\ref{lub''})
with $\alpha\rightarrow\infty$, i.e. with $(2\alpha-1)/(3\alpha+1)\rightarrow2/3$
in the first term in (\ref{lub''}). In other words, the dimensionless
zero-toughness solution, $\Omega_{m}(\xi)$ and $\gamma_{m}$, for
exponential injection is given by that for the power-law case in the
$\alpha\rightarrow\infty$ limit\footnote{This equivalence is for the non-dimensional solutions only. Corresponding
dimensional solutions are distinct, as they are recovered from the
non-dimensional ones by applying the distinct sets of time-dependent
scales $W_{m}$ and $L_{m}$, given by (\ref{exp}) and (\ref{power'})
for exponential and power-law injections, respectively.}.

\subsubsection*{Particular Zero-Toughness Solutions for $\alpha=0$ and $\alpha=4/3$}

Differential equation (\ref{lub''}) has an analytical zero-toughness
solution for specific values of $\alpha$, that is when $\alpha=0$
in the fixed-volume-case, and when $\alpha=4/3$, which is a particular
injection scenario corresponding to a constant rate of crack growth
($\overline{L}_{m}\propto t$), respectively. These two solutions
are: 
\begin{equation}
\text{fixed-volume, }\alpha=0:\quad\Omega_{m}=\left(\frac{1-\xi^{2}}{2}\right)^{1/3},\quad\vartheta_{m}=\xi,\quad\gamma_{m}=1.49757,\label{0}
\end{equation}
\begin{equation}
\text{fixed-propagation-rate,}\,\alpha=4/3:\quad\Omega_{m}=(1-\xi)^{1/3},\quad\vartheta_{m}=1,\quad\gamma_{m}=4/3,\label{4/3}
\end{equation}
where $\vartheta(\xi)$ and $\gamma$ were evaluated from (\ref{Mtip})
and (\ref{global'}), respectively.

\subsubsection*{General Zero-Toughness Solution}

General zero-toughness solutions for arbitrary injection power-law
can be obtained by series expansion. As already pointed out, the original
PKN formulation corresponds to the zero-toughness limit of the model
suggested here. The first treatment of this case is due to \citet{Nord72},
and its complete solution for a constant fluid injection rate ($\alpha=1$)
was first formulated by \citet{Kemp90}. 

Kemp's solution can be generalized to arbitrary injection power law
$V\propto t^{\alpha}$ with $\alpha>0$ as 
\begin{equation}
\Omega_{m}=(1-\xi)^{1/3}\sum_{j=0}^{\infty}A_{j}(\alpha)(1-\xi)^{j},\qquad\gamma_{m}^{-1}=\int_{0}^{1}\Omega_{m}d\xi=\frac{3}{4}\sum_{j=0}^{\infty}\frac{A_{j}(\alpha)}{1+3j/4}\label{inj}
\end{equation}
Coefficients $A_{j}(\alpha)$ are obtained by substituting the opening
series into lubrication equation (\ref{lub''}) and expanding the
result near the tip, $\xi=1$,
\[
A_{0}=1,\quad A_{1}=\frac{3\alpha/4-1}{6(3\alpha+1)},\quad A_{2}=\frac{(3\alpha/4-1)(1-51\alpha/28)}{6^{2}(3\alpha+1)^{2}},...\frac{}{}
\]
where $j=0$ term provides the tip asymptote (\ref{Ktip}). It is
evident that $A_{j\ge1}\propto3\alpha/4-1$, thus, validating that
$j=0$ term corresponds to the exact solution in the particular case
of $\alpha=4/3$, see (\ref{4/3}).

\begin{figure}[H]
\begin{centering}
\includegraphics[scale=0.5]{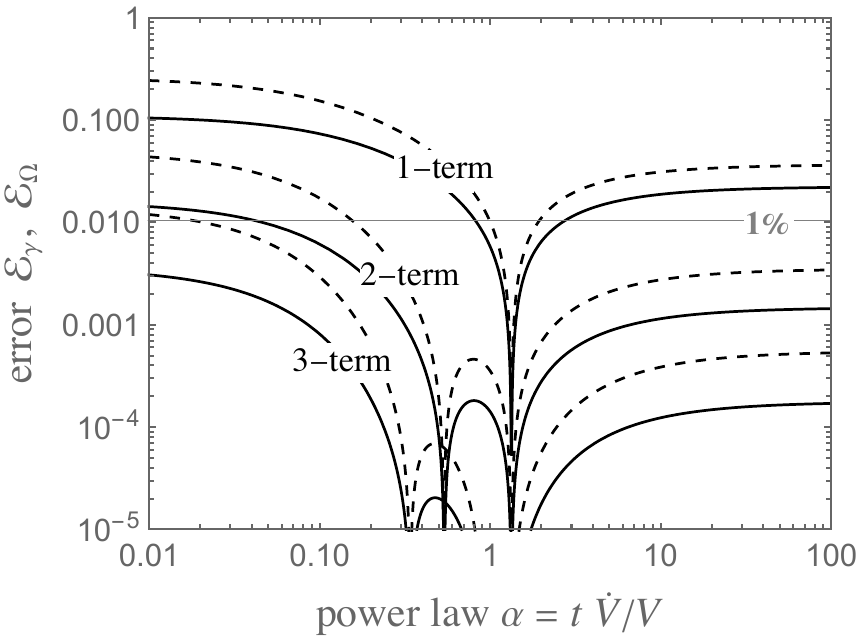}
\par\end{centering}
\caption{Power-law injection, zero-toughness. Relative error for the fracture
half-length (solid) and opening at the inlet (dashed) of the series
solution (\ref{inj}) when truncated to one, two, and three terms,
respectively, as a function of injection power-law $\alpha$. \label{fig:conv}}
\end{figure}

Examination of the series convergence (Fig. \ref{fig:conv}) shows
that the single-term-solution, i.e. $\Omega_{m}\approx(1-\xi)^{1/3}$
and $\gamma_{m}^{-1}\approx3/4$, provides a very good approximation
for the half-length and opening (within $1\%$) for injections at
a non-decreasing rate in time, $\alpha\ge1$, while approximation
becomes progressively less adequate for decreasing value of $\alpha<1$.
Furthermore, Fig. \ref{fig:conv} shows that the two-term-solution,
$\Omega_{m}\approx(1-\xi)^{1/3}+A_{1}(\alpha)(1-\xi)^{4/3}$ and $\gamma_{m}^{-1}\approx3/4+3A_{1}(\alpha)/7$,
gives the half-length within $1\%$ for the entire range of injection
exponent $\alpha\ge0$.  
\begin{figure}[H]
\begin{centering}
(a)\includegraphics[scale=0.45]{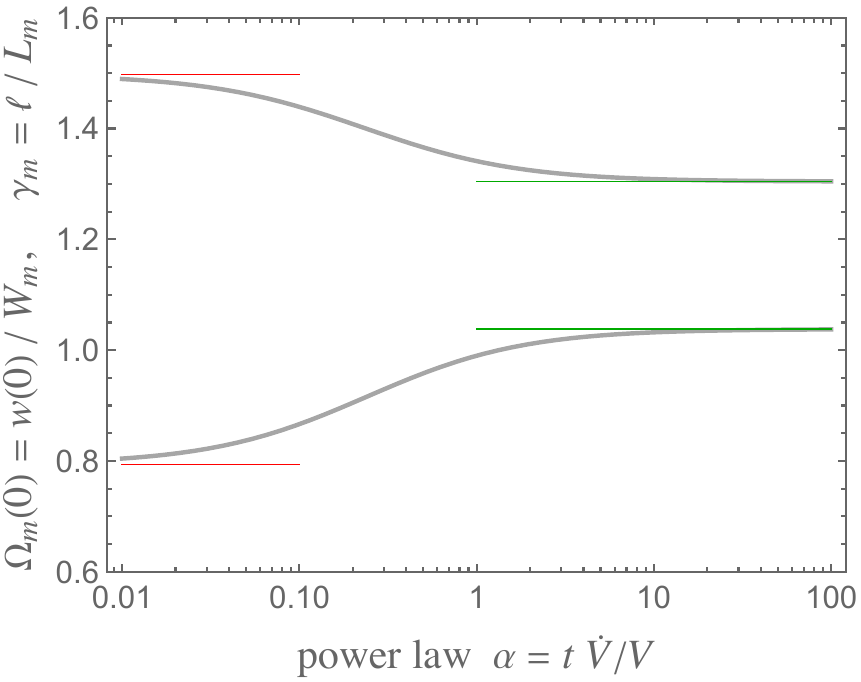}\quad{}(b)\includegraphics[scale=0.45]{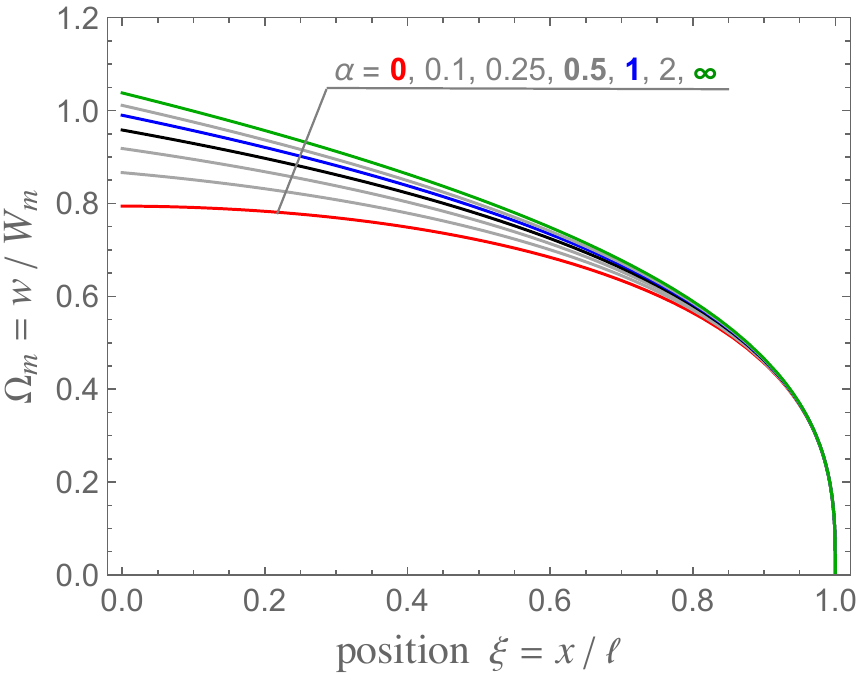}
\par\end{centering}
\caption{Power-law injection, zero-toughness solution in the \emph{implicit
viscosity scaling}, (\ref{inj}). (\textbf{a}) Normalized opening/net-pressure
at the inlet $\Omega_{m}(0)=w(0)/W_{m}=p(0)/(\bar{E}W_{m}/b)$ and
crack half-length $\gamma_{m}=\ell/L_{m}$ as a function of injection
power-law $\alpha$, $V(t)\propto t^{\alpha}$. The fixed-volume $\alpha=0$
and equivalent exponential $\alpha=\infty$ limits are indicated by
thin lines. (\textbf{b}) Distribution of the normalized opening/net-pressure
$\Omega_{m}=w/W_{m}$ along the crack as a function of normalized
position $\xi=x/\ell$, from the inlet $\xi=0$ to the tip $\xi=1$,
for various values of $\alpha$. \label{fig:length}}
\end{figure}
\begin{figure}[H]
\begin{centering}
(a)\includegraphics[scale=0.17]{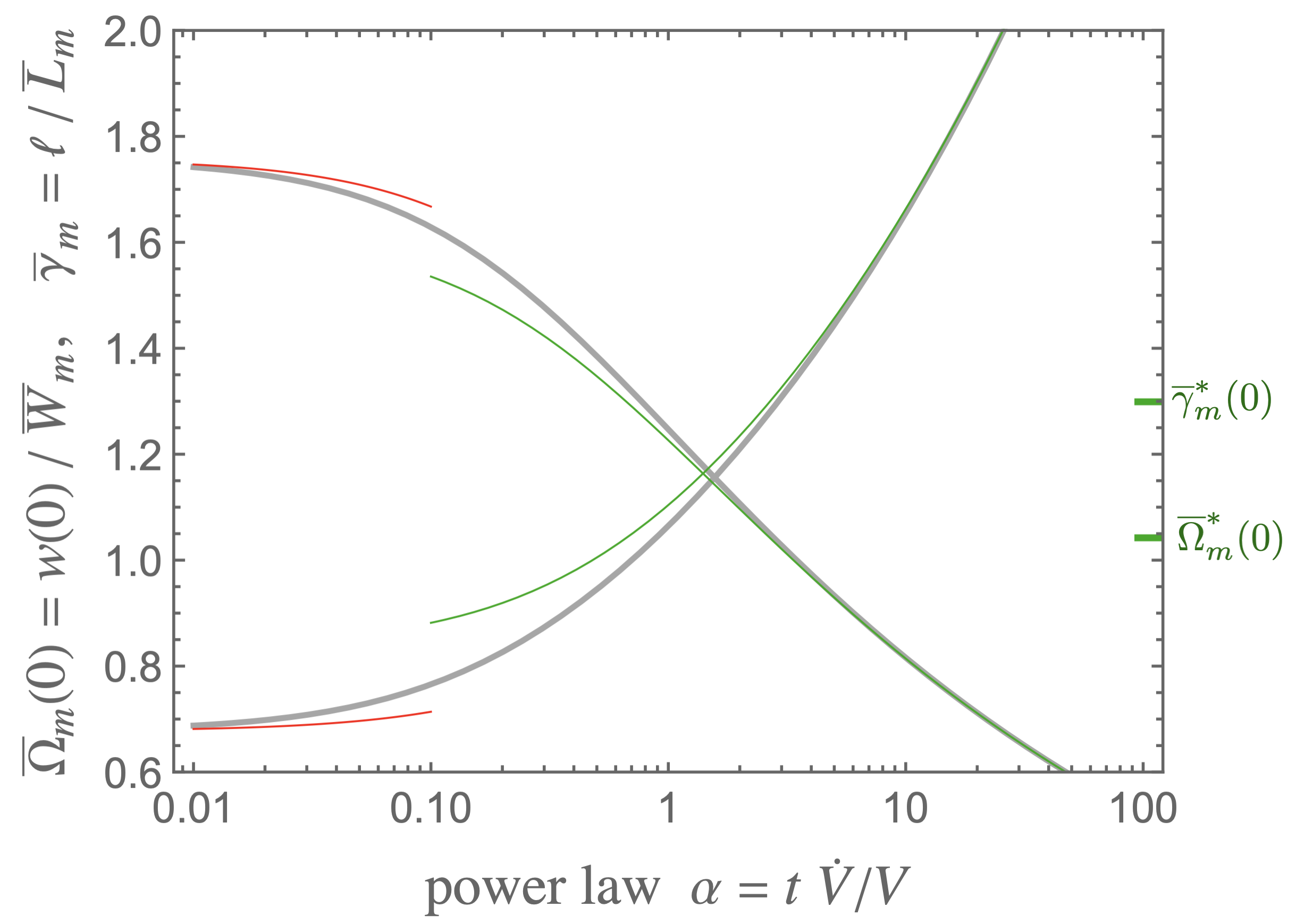}\quad{}(b)\includegraphics[scale=0.46]{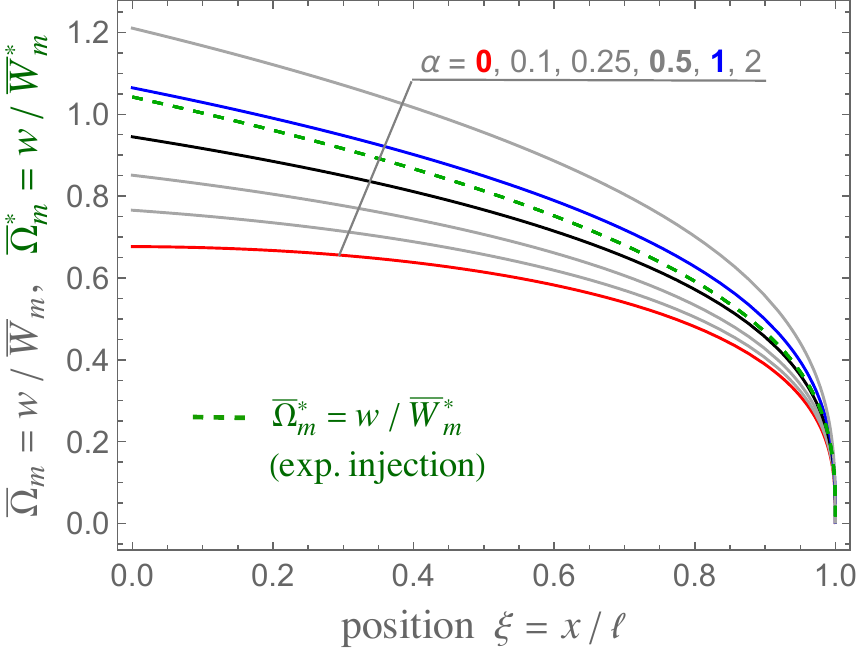}
\par\end{centering}
\caption{Power-law injection, zero-toughness solution (Fig. \ref{fig:length})
rescaled to the \emph{explicit viscosity scaling}. (\textbf{a}) Normalized
opening/net-pressure at the inlet $\bar{\Omega}_{m}(0)=w(0)/\overline{W}_{m}=p(0)/(\bar{E}\overline{W}_{m}/h)$
and crack half-length $\bar{\gamma}_{m}=\ell/\overline{L}_{m}$ as
a function of injection power-law $\alpha$, $V(t)\propto t^{\alpha}$.
Small and large $\alpha$ asymptotes are shown by thin lines. (\textbf{b})
Distribution of the normalized opening/net-pressure along the crack,
$\bar{\Omega}_{m}=w/\overline{W}_{m}$ for a power-law $0\le\alpha<\infty$
and $\bar{\Omega}_{m}^{*}=w/\overline{W}_{m}^{*}$ for an exponential
injection, as a function of normalized position $\xi=x/\ell$. Dependence
of the solution on time follows from that of the scales: $\overline{W}_{m}(t)=(3\bar{\mu}V^{2}(t)/4\bar{E}b\,t)^{1/5}\propto t^{(2\alpha-1)/5}$
and $\overline{L}_{m}(t)=V(t)/(2b\overline{W}_{m}(t))\propto t^{(3\alpha+1)/5}$
for a power-law, and $\overline{W}_{m}^{*}(t)=\overline{W}_{m}(t_{*})\,e^{2(t/t_{*}-1)/5}$
and $\overline{L}_{m}^{*}(t)=\overline{L}_{m}(t_{*})\,e^{3(t/t_{*}-1)/5}$
for an exponential-law injection.  \label{fig:length-explicit}}
\end{figure}

Zero-toughness solution in the entire range of power-law exponent
$\alpha$, $0\le\alpha\le\infty$, is shown in Fig. \ref{fig:length}.
Fig. \ref{fig:length}a shows the normalized crack length $\gamma_{m}$
and opening at the inlet $\Omega_{m}(0)$ as a function of the power-law
$\alpha$. Spatial distribution of the normalized fracture opening
is shown on Fig. \ref{fig:length}b for various values of $\alpha$. 

Normalized zero-toughness solution, (\ref{inj}) and Fig. \ref{fig:length},
is expressed in terms of implicit scales $L_{m}$ and $W_{m}$, (\ref{W_m}).
Corresponding normalized solution using explicit scales $\overline{L}_{m}$
and $\overline{W}_{m}$, (\ref{scaling-m-bar}), and, therefore, subsequently
corresponding \emph{dimensional solution} can be readily obtained
from relation (\ref{power'}) for power-law injection ($0\le\alpha<\infty$),
and from similar relation (\ref{exp}-\ref{scales_exp}) for exponential
injection ($\alpha=\infty$). For example, for the explicit dimensionless
half-length $\overline{\gamma}_{m}=\ell/\overline{L}_{m}=(5/(3\alpha+1))^{1/5}\gamma_{m}^{3/5}$
 for power-law and $\overline{\gamma}_{m}=\ell/\overline{L}_{m}^{*}=(5/3)^{1/5}\gamma_{m}^{3/5}$
 for exponential injection. Resulting explicit dimensionless zero-toughness
solution is plotted on Fig. \ref{fig:length-explicit}.

Table \ref{tab:zero} lists numerical values accurate to at least
six significant digits of the normalized solution for the fracture
half-length and opening at the inlet, in both the implicit and explicit
scales, for various values of injection power-law exponent. We particularly
highlight solutions for practically-relevant injection scenarios corresponding
to injections with fixed volume ($\alpha=0$), constant injection
pressure ($\alpha=1/2$), constant volume-rate ($\alpha=1$), and
exponentially increasing volume, rate thereof, and injection pressure
($\alpha=\infty$).

\begin{table}[th]
\centering{}{\footnotesize{}\caption{{\footnotesize{}Zero-toughness solution for the dimensionless fracture
half-length and opening/net-pressure at the inlet using implicit,
$\gamma_{m}=\ell/L_{m}$ and $\Omega_{m}(0)=w(0)/W_{m}=p(0)/(\bar{E}W_{m}/b)$,
and explicit, $\overline{\gamma}_{m}=\ell/\overline{L}_{m}$ and $\overline{\Omega}_{m}(0)=w(0)/\overline{W}_{m}=p(0)/(\bar{E}\overline{W}_{m}/b)$,
scales for various values of injection power-law $\alpha$ ($V\propto t^{\alpha}$).
For exponential injection ($V\propto e^{t/t_{*}}$), the implicit
dimensionless solution is given by that for the power-law one with
$\alpha\rightarrow\infty$, while the explicit dimensionless solution
is using amended explicit length and opening scales, $\overline{\gamma}_{m}=\ell/\overline{L}_{m}^{*}$
and $\overline{\Omega}_{m}(0)=w(0)/\overline{W}_{m}^{*}=p(0)/(\bar{E}\overline{W}_{m}^{*}/b)$.}
{\footnotesize{}The two sets of length and opening scales are related
via (\ref{power'}) for power-law injection, and by (\ref{exp}-\ref{scales_exp})
for exponential injection.}\label{tab:zero}}
}%
\begin{tabular}{llllll}
\toprule 
\multirow{2}{*}{{\footnotesize{}Power-law $\alpha$}} & \multirow{2}{*}{{\footnotesize{}Description}} & \multicolumn{2}{l}{{\footnotesize{}Implicit}} & \multicolumn{2}{l}{{\footnotesize{}Explicit}}\tabularnewline
\cmidrule{3-6} \cmidrule{4-6} \cmidrule{5-6} \cmidrule{6-6} 
 &  & {\footnotesize{}$\gamma_{m}$} & {\footnotesize{}$\Omega_{m}(0)$} & {\footnotesize{}$\overline{\gamma}_{m}$} & {\footnotesize{}$\overline{\Omega}_{m}(0)$}\tabularnewline
\midrule
\textbf{\footnotesize{}0} & {\footnotesize{}fixed injection volume} & {\footnotesize{}1.49757} & {\footnotesize{}0.79370} & {\footnotesize{}1.75815} & {\footnotesize{}0.67607}\tabularnewline
{\footnotesize{}0.1} &  & {\footnotesize{}1.43910} & {\footnotesize{}0.86642} & {\footnotesize{}1.62877} & {\footnotesize{}0.76553}\tabularnewline
{\footnotesize{}0.25} &  & {\footnotesize{}1.39748} & {\footnotesize{}0.91836} & {\footnotesize{}1.50797} & {\footnotesize{}0.85107}\tabularnewline
\textbf{\footnotesize{}1/2} & {\footnotesize{}fixed injection pressure} & {\footnotesize{}1.36597} & {\footnotesize{}0.95815} & {\footnotesize{}1.38506} & {\footnotesize{}0.94494}\tabularnewline
\textbf{\footnotesize{}1} & {\footnotesize{}fixed injection volume rate} & {\footnotesize{}1.34113} & {\footnotesize{}0.98993} & {\footnotesize{}1.24699} & {\footnotesize{}1.06466}\tabularnewline
{\footnotesize{}4/3} & {\footnotesize{}fixed crack propagation rate} & {\footnotesize{}1.33333} & {\footnotesize{}1.00000} & {\footnotesize{}1.18840} & {\footnotesize{}1.12196}\tabularnewline
{\footnotesize{}2} &  & {\footnotesize{}1.32469} & {\footnotesize{}1.01121} & {\footnotesize{}1.10674} & {\footnotesize{}1.21036}\tabularnewline
\midrule
$\boldsymbol{\infty}$ & {\footnotesize{}exponential injection} & {\footnotesize{}1.30424} & {\footnotesize{}1.03802} & {\footnotesize{}1.29893} & {\footnotesize{}1.04227}\tabularnewline
\bottomrule
\end{tabular}
\end{table}

\section{Transient Solutions}

When considering a fracture propagating in the rock with finite toughness,
the solution is bound to evolve in time between the two limiting propagation
regimes discussed in Section \ref{Sec:regimes}. More specifically,
given the dependence of the normalized viscosity on time, $\mathcal{M}\propto V^{2}(t)/t$,
this evolution will take place from the initially toughness-dominated
to the eventually viscosity-dominated regime when injection is fast
enough (injected volume $V(t)$ increases faster than $\sqrt{t}$),
and the reversed evolution, from the viscosity- towards the toughness-
domination with time, will take place when the injection is slow enough
($V(t)$ increases slower than $\sqrt{t}$). The former is true, for
example, for injection with fixed rate, $\mathcal{M}\propto t$, while
the latter, e.g., for injection with fixed volume, $\mathcal{M}\propto1/t$.
The threshold injection law, $V\propto\sqrt{t}$ and $\mathcal{M}=\text{const}$,
which separates the two types of evolutionary behavior, corresponds
to injection with fixed source pressure.

A numerical solution of the fracture evolution in space-time to any
desired degree of accuracy can be obtained by means of the method-of-lines,
which relies on a discretization of the governing equations in space
and then solving resulting system of ODEs at the spatial nodes continuously
in time (see Appendix \ref{App:num} for details). As an alternative
to this fully-numerical solution, we also propose an approximate \emph{Equation-of-Motion}
(EofM) for the elongated hydraulic fracture - a single ODE in time
which allows for a simplified, accurate and very expedient approximate
solution.

In this Section, we formulate the EofM approach first and then provide
both the fully-numerical and the EofM transient solutions for (i)
power-law-injection, and (ii) injection at a constant rate followed
by the shut-in.

\subsection{Approximate Equation-of-Motion}

To facilitate a simplified solution approach, we choose to approximate
the crack opening spatiotemporal evolution by the form suggested by
the small-viscosity solution, given in the \emph{toughness scaling}
by the firsts in (\ref{M1_inj}) and (\ref{M1_shut}), respectively,
and repeated below together with the resulting approximation for the
normalized fluid velocity $\vartheta_{k}=-\partial\Omega_{k}^{3}/\partial\xi$,
\begin{equation}
\text{injection, \ensuremath{\dot{V}>0}:}\quad\Omega_{k}(\mathcal{\mathbb{M}},\xi)\approx\left(1+\mathcal{\mathbb{M}}\,(1-\xi)\right)^{1/3},\qquad\vartheta_{k}(\xi)\approx1\label{inj_approx}
\end{equation}
\begin{equation}
\text{fixed-volume, \ensuremath{\dot{V}=0}:}\quad\Omega_{k}(\mathcal{\mathbb{M}},\xi)\approx\left(1+\mathbb{M}\,\frac{2}{\pi}\cos\frac{\pi\xi}{2}\right)^{1/3},\qquad\vartheta_{k}(\xi)\approx\sin\frac{\pi\xi}{2}\label{shut-in_approx}
\end{equation}
Time-dependence of the approximate solution follows solely from that
of the normalized, implicit viscosity parameter $\mathcal{\mathbb{M}}(t)$.
Contrary to the small-viscosity asymptote, here $\mathcal{\mathbb{M}}(t)$
is not necessarily small and is an unknown (part of the solution).
We note in passing, that the chosen opening approximation can also
be framed as an approximate continuation of the tip asymptote, the
second in (\ref{Ktip}), to the remainder of the crack, that is respecting
the inlet (flow/no-flow) boundary condition. Latter approach was originally
used in \citep{Garagash19,Dontsov16,Dontsov17,garagash2022-HF-notes}
to formulate EofM for plane-strain and radial hydraulic fractures. 

Corresponding approximation for the normalized crack length, $\gamma_{k}=\ell/L_{k}$,
follows by evaluating the global fluid continuity (\ref{global'})
for the assumed form for the opening, (\ref{inj_approx})-(\ref{shut-in_approx}),
\begin{equation}
\gamma_{k}\approx\varUpsilon(\mathbb{M})=\left(\int_{0}^{1}\Omega_{k}(\mathcal{\mathbb{M}},\xi)d\xi\right)^{-1}\label{omega}
\end{equation}
where
\begin{align*}
\text{injection, \ensuremath{\dot{V}>0}:}\quad & \varUpsilon(\mathbb{M})=\frac{4}{3}\frac{\mathbb{M}}{(1+\mathbb{M})^{4/3}-1}\\
\text{fixed-volume, \ensuremath{\dot{V}=0}:}\quad & \varUpsilon(\mathbb{M})=\left[\,_{2}F_{1}\left(\left\{ \begin{array}{c}
-1/6\\
1/3
\end{array}\right\} ,1,\frac{4\mathbb{M}^{2}}{\pi^{2}}\right)+\frac{4\mathbb{M}}{3\pi^{2}}\,_{3}F_{2}\left(\left\{ \begin{array}{c}
1/3\\
5/6\\
1
\end{array}\right\} ,\left\{ \begin{array}{c}
3/2\\
3/2
\end{array}\right\} ,\frac{4\mathbb{M}^{2}}{\pi^{2}}\right)\right]^{-1}
\end{align*}
and $\,_{2}F_{1}$ and $\,_{3}F_{2}$ are generalized hypergeometric
functions \citep{AbSt72}.

Finally, an ODE governing evolution of $\mathcal{\mathbb{M}}$ with
time is obtained from the 2nd expression in (\ref{K}), recast as
$t\dot{\ell}/\ell=\mathcal{\mathbb{M}}/(\gamma_{k}^{2}\,\mathcal{M})$,
and upon substituting $t\dot{\ell}/\ell=t\dot{V}/V+t\dot{\gamma}_{k}/\gamma_{k}$,
see (\ref{L'}) in the toughness scaling, and $\gamma_{k}=\varUpsilon(\mathbb{M})$,
see (\ref{omega}), 
\begin{equation}
\frac{t\dot{V}}{V}+\frac{d\ln\varUpsilon(\mathbb{M})}{d\ln t}=\frac{\mathcal{\mathbb{M}}}{\mathcal{M}(t)\,\varUpsilon^{2}(\mathbb{M})}\label{ODE}
\end{equation}
where explicit dimensionless viscosity parameter $\mathcal{M}(t)$
is given by the 2nd in (\ref{K_exp}). Further, it may prove convenient
to reformulate this ODE using $\mathcal{M}$ as the `time' variable.
Relating the log-derivatives, $d\ln\mathcal{M}/d\ln t=2t\dot{V}/V-1$,
we have 
\begin{equation}
\frac{t\dot{V}}{V}+\left(2\frac{t\dot{V}}{V}-1\right)\frac{d\ln\varUpsilon(\mathbb{M})}{d\ln\mathcal{M}}=\frac{\mathcal{\mathbb{M}}}{\mathcal{M}\,\varUpsilon^{2}(\mathbb{M})}\label{ODE'}
\end{equation}

Evidently, EofM in this form is particularly useful when injected
fluid volume is a power-law of time, i.e. when $\alpha=t\dot{V}/V=\text{const}$,
and solution $\mathcal{\mathbb{M}}=\mathcal{\mathbb{M}}(\mathcal{M}$)
of (\ref{ODE'}) depends on the single parameter - the fluid-volume-power-law-exponent
$\alpha$. Once this ODE is solved, the approximate solution for the
normalized half-length, (\ref{omega}), and opening, (\ref{inj_approx})
when $\alpha>0$ and (\ref{shut-in_approx}) when $\alpha=0$, is
complete.

\subsection{Initial Conditions}

Initial conditions at small fracture propagation time for either full
numerical treatment of the problem (Appendix \ref{App:num}) or in
the approximate EofM approach discussed in the above normally correspond
to small-$\mathcal{M}$ asymptote for the `fast', $\alpha=t\dot{V}/V>1/2$,
and large-$\mathcal{M}$ asymptote for the `slow', $\alpha<1/2$,
injection. 

Small-$\mathcal{M}$ asymptotic solution in the toughness-scaling
is given by (\ref{M1_inj}), which we re-write here retaining the
leading-order term only for simplicity
\begin{equation}
\text{\ensuremath{\mathcal{M}\ll1:\quad}}\mathbb{M}\approx\alpha\mathcal{M},\quad\gamma_{k}\approx1,\quad\text{\ensuremath{\Omega_{k}\approx1}}\label{small-M}
\end{equation}

Large-$\mathcal{M}$ asymptotic solution corresponds to the zero-toughness
solution given in the viscosity scaling by (\ref{inj}). Converting
the latter to the toughness-scaling 
\begin{equation}
\mathcal{M}\gg1:\quad\mathbb{M}=\left(\frac{(3\alpha+1)\gamma_{m}^{2}}{5}\right)^{3/5}\mathcal{M}^{3/5},\quad\gamma_{k}=\mathbb{M}^{-1/3}\gamma_{m},\quad\Omega_{k}=\mathbb{M}^{1/3}\Omega_{m}\label{large-M}
\end{equation}
where $\gamma_{m}=\gamma_{m}(\alpha)$ and $\Omega_{m}=\Omega_{m}(\xi;\alpha)$
is the zero-toughness solution (\ref{inj}).

\subsection{Solution for Power-Law Fluid Injection\label{Sec:power}}

This section presents regime-transient solution for continuous power-law
injection $V\propto t^{\alpha}$ with non-zero material toughness.
Zero-viscosity (toughness-dominated regime) and zero-toughness (viscosity-dominated
regime) end-members of this solution have been given in Sections \ref{Sec:inf}
and \ref{Sec:zero}. 
\begin{figure}[H]
\begin{centering}
(a)\includegraphics[scale=0.47]{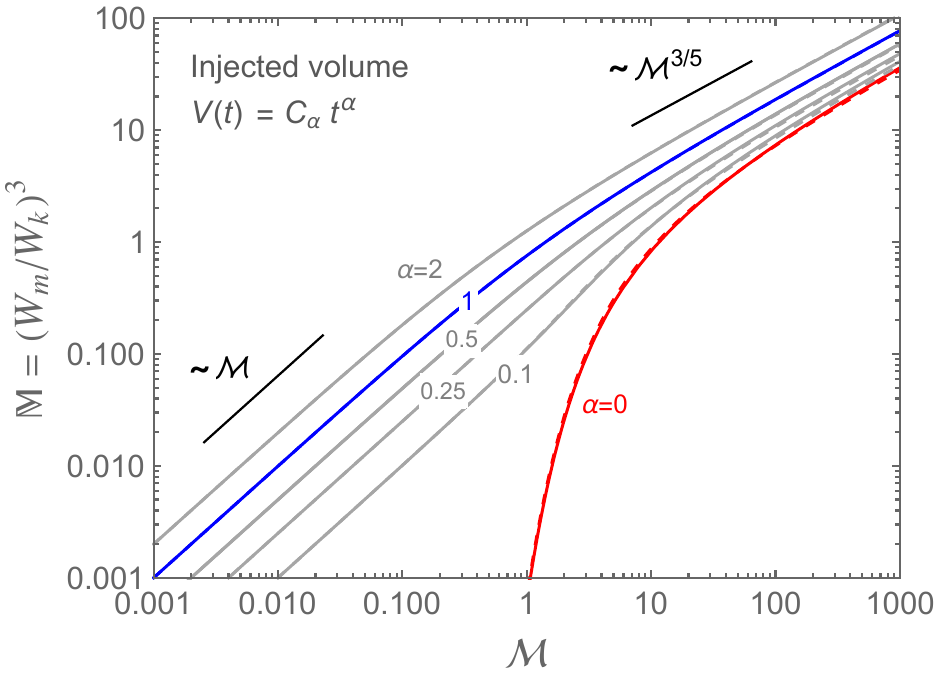}
\par\end{centering}
\begin{centering}
(b)\includegraphics[scale=0.45]{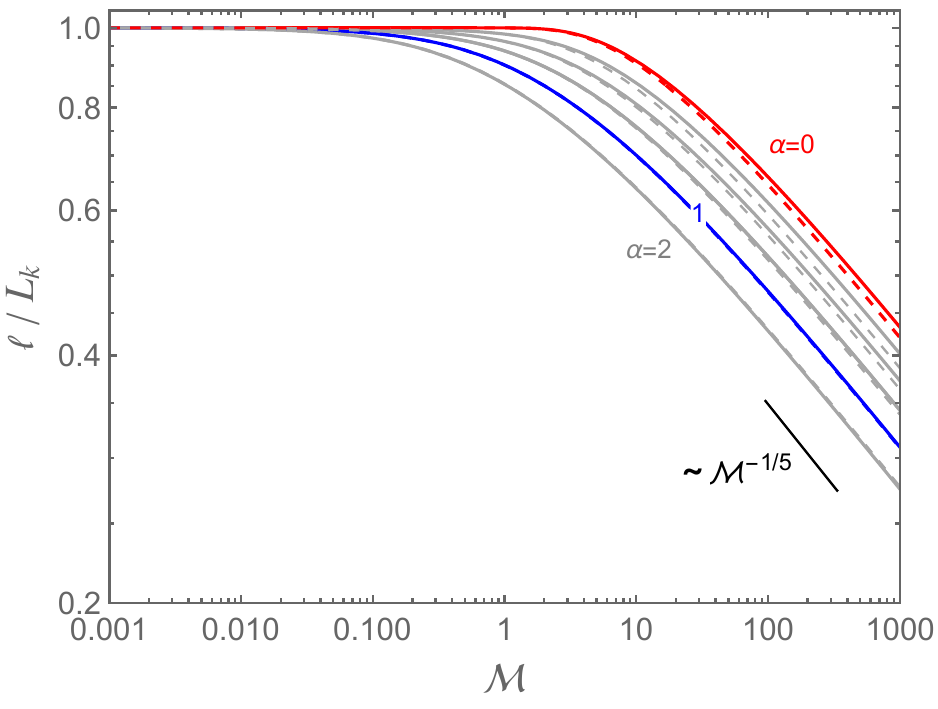}
\par\end{centering}
\begin{centering}
(c)\includegraphics[scale=0.44]{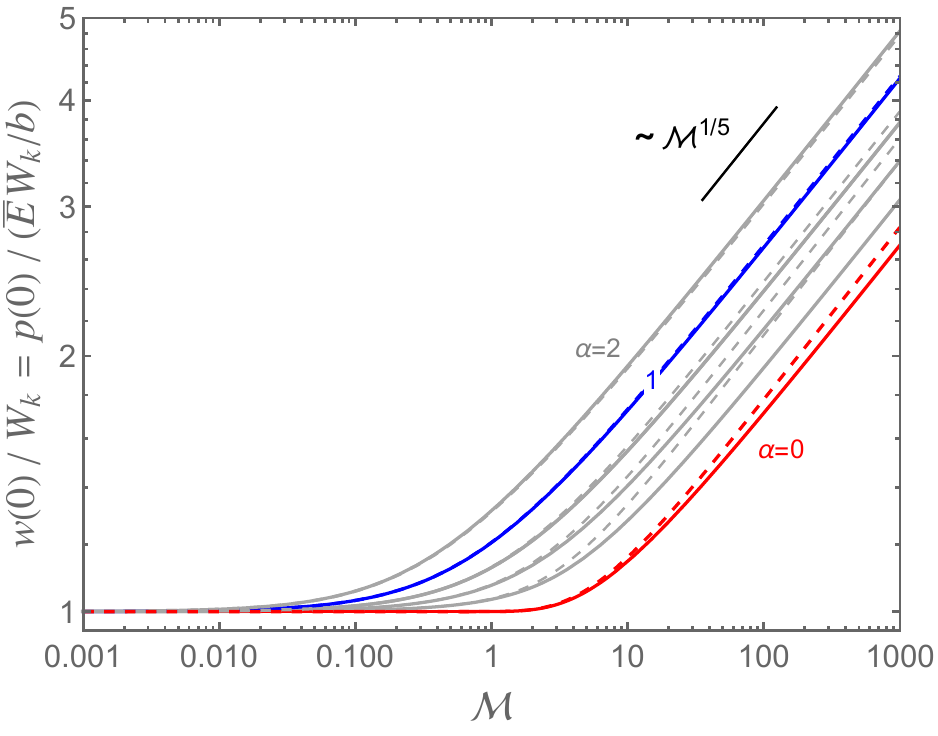}
\par\end{centering}
\caption{Power-law injection solution in the toughness scaling. Equation-of-Motion
(dashed) and the fully numerical Method-of-Lines (solid) solutions
for hydraulic fracture propagation driven by a power-law injection
$V(t)=Ct^{\alpha}$ for various values of $\alpha$: $0$ (constant
volume), 0.1, 0.25, 0.5 (constant injection pressure), 1 (constant
rate of injection), and 2. \textbf{(a)} Implicit dimensionless viscosity
parameter $\mathbb{M}\propto\ell\dot{\ell}$ evolution with the explicit
dimensionless viscosity parameter $\mathcal{M}(t)\propto V^{2}(t)/t$.
\textbf{(b,c)} Corresponding evolution of the normalized crack half-length
$\ell/L_{k}(t)$ and opening/net-pressure at the inlet $w(0)/W_{k}=p(0)/(\bar{E}W_{k}/b)$
with $\mathcal{M}(t)$. Dependence of the solution on time follows
from that of the scales, $W_{k}=\bar{K}\sqrt{b}/\bar{E}$ and $L_{k}(t)=V(t)/(2bW_{k})$,
and of the evolution parameter $\mathcal{M}(t)=(3\bar{\mu}\bar{E}^{4}/4\bar{K}^{5}b^{7/2})(V^{2}(t)/t)$,
as further explored in Fig. \ref{fig:power'}. \label{fig:power}}
\end{figure}
 Fig. \ref{fig:power} shows the dimensionless solution in the \emph{toughness-scaling}
for various values of injection power-law $\alpha$, namely the evolution
of dimensionless implicit viscosity parameter $\mathbb{M}$, the normalized
fracture half-length, and opening at the inlet with explicit, time-dependent
dimensionless viscosity parameter $\mathcal{M}(t)$. The solution
corresponds to the transition from the toughness- to viscosity- dominated
regime of propagation with increasing value of $\mathcal{M}$, with
corresponding asymptotic solutions given by the zero-viscosity and
zero-toughness solutions discussed earlier, respectively. Given different
sense of time-dependence of $\mathcal{M}$ depending on the power-law
$\alpha$, $\mathcal{M}\propto t^{2\alpha-1}$, the aforementioned
regime transition in time is from the toughness- to the viscosity-
regime for `fast' injections $\alpha>1/2$ and is reversed for `slow'
injections $0\le\alpha<1/2$. Change of the archetype of the transient
solution takes place for an injection with $\alpha=1/2$, which correspond
to injection with constant injection pressure at the crack inlet and
corresponding self-similar solution ($\mathcal{M}$ is time-independent
constant). Thus, `fast' injections leading to toughness-to-viscosity
regime transition in time are characterized by increasing fluid pressure
(and therefore crack opening) at the inlet with time, while `slow'
injections leading to the viscosity-to-toughness transition correspond
to decreasing injection pressure in time. 

\subsubsection*{Explicit time-dependence of the solution - transitional $MK$ scaling}

Given non-trivial time-dependence of both the evolution parameter
$\mathcal{M}(t)$ and lengthscale $L_{k}(t)$ pertaining to the toughness-scaling
of the solution in Fig. \ref{fig:power}, we seek to rescale into
a different, time-independent scaling in order to illustrate the explicit
dependence of the solution for the crack opening $w$ (net fluid pressure
$p=E'\,w/b$) and the crack half-length $\ell$ on time. 

\begin{figure}[H]
\begin{centering}
(a)\includegraphics[scale=0.45]{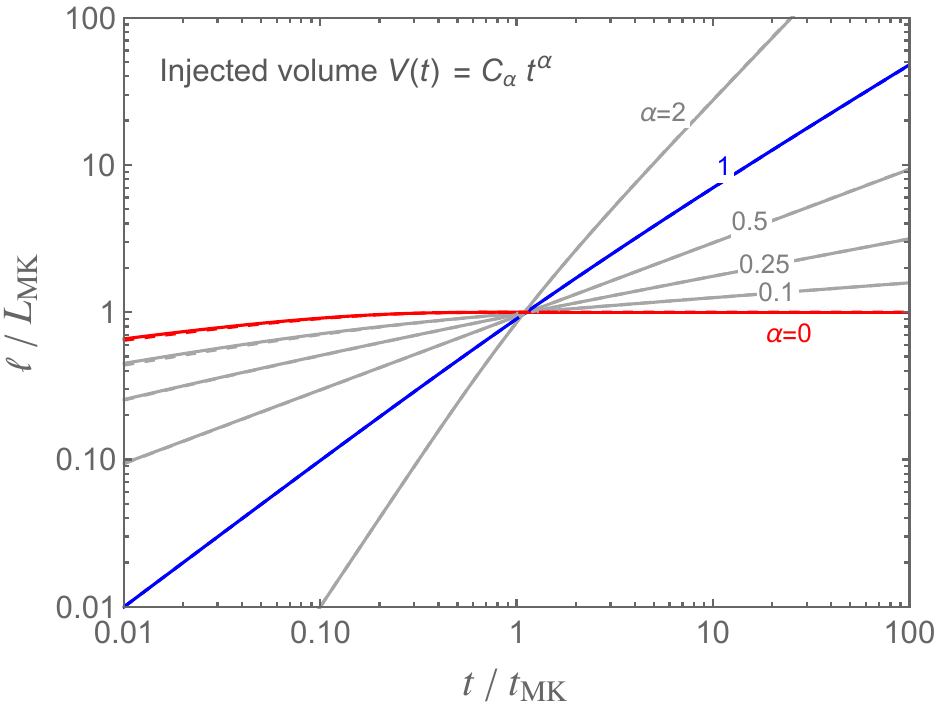}\quad{}(b)\includegraphics[scale=0.44]{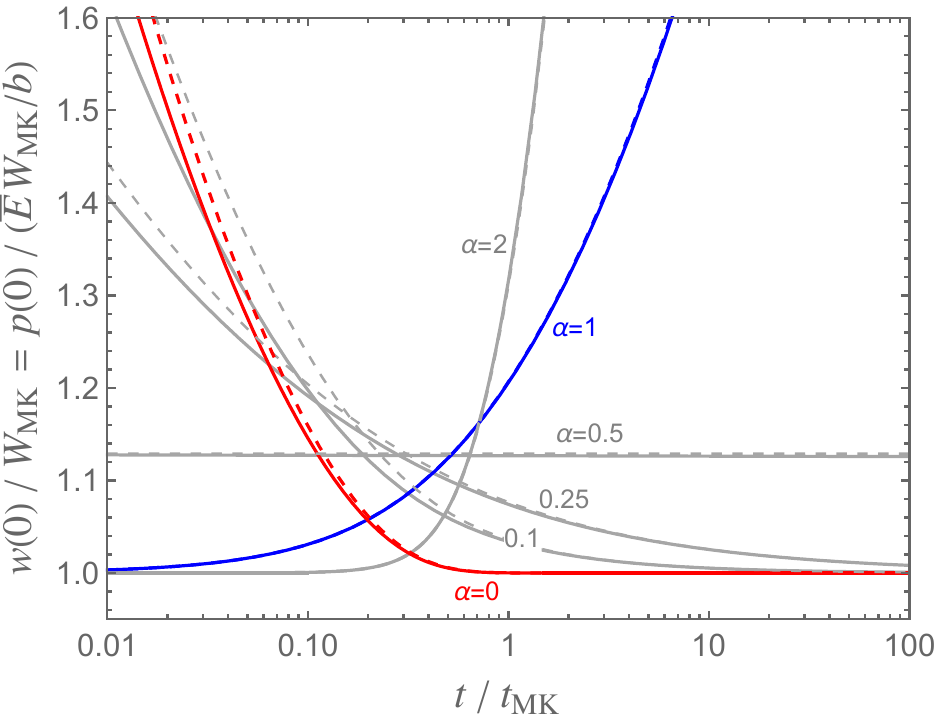}
\par\end{centering}
\caption{Power-law injection solution in the transitional $MK$ scaling (explicit
time dependence). Solution for the crack half-length and opening/net-pressure
of Fig. \ref{fig:power}(b-c) re-scaled using the time-independent
scales, $\ell/L_{MK}(\alpha)$ and $w(0)/W_{MK}=p(0)/(\bar{E}W_{MK}/h)$,
and shown as a function of normalized time $\tau=t/t_{MK}(\alpha)$.
The transitional scales are $W_{MK}=W_{k}$ and $t_{MK}(\alpha)$,
$L_{MK}(\alpha)$ given by (\ref{scales_MK}), while evolution parameter
of Fig. \ref{fig:power} is $\mathcal{M}(t)=\tau^{2\alpha-1}$. Note
that `fast' $\alpha>0.5$ injection corresponds to the fracture evolution
in time from the toughness to the viscosity-dominated regime, while
the opposite is true for `slow' $\alpha<0.5$ injection, including
the constant-volume case $\alpha=0$. \label{fig:power'}}
\end{figure}

Let us define transitional timescale as the time $t=t_{MK}$ when
the viscosity and toughness crack aperture scales assume the same
value, further denoted as transitional opening scale $W_{MK}$:

\[
W_{MK}=\overline{W}_{m}(t_{MK})=W_{k}\quad\Rightarrow\quad\frac{V^{2}(t_{MK})}{t_{MK}}=\frac{4}{3}\frac{\bar{K}^{5}b^{7/2}}{\bar{\mu}\bar{E}^{4}}
\]
By their definitions, the dimensionless, time-dependent viscosity
and toughness parameters attain value of unity at $t=t_{MK}$, i.e.
$\mathcal{M}(t_{MK})=\mathcal{K}(t_{MK})=1$.

Corresponding transitional crack lengthscale is then defined by
\[
L_{MK}\equiv\overline{L}_{m}(t_{MK})=L_{k}(t_{MK})
\]

Intuitively, $t_{MK}$, $L_{MK}$, and $W_{MK}$ should scale the
time, crack length and opening when the transition between the viscosity
and toughness dominated regimes takes place.

For power-law injection $V(t)=C_{\alpha}t^{\alpha}$ (such that $C_{0}=V_{0}$
for fixed volume injection, and $C_{1}=Q_{o}$ for constant rate injection)
we have
\begin{equation}
t_{MK}=\left(\frac{3}{4}\frac{\bar{\mu}\bar{E}^{4}C_{\alpha}^{2}}{\bar{K}^{5}b^{7/2}}\right)^{1/(1-2\alpha)},\qquad L_{MK}=\left(\frac{3^{\alpha}}{2}\frac{\bar{\mu}^{\alpha}\bar{E}^{1+2\alpha}C_{\alpha}}{\bar{K}^{1+3\alpha}b^{(3+\alpha)/2}}\right)^{1/(1-2\alpha)}\label{scales_MK}
\end{equation}
For example, taking the fixed volume ($\alpha=0$, $C_{0}=V_{0}$)
and fixed injection rate ($\alpha=1$, $C_{1}=Q_{0}$) cases, the
corresponding expressions for the time and lengthscales evaluate to:
\begin{align}
\alpha & =0:\quad t_{MK}=\frac{3}{4}\frac{\bar{\mu}\bar{E}^{4}V_{0}^{2}}{\bar{K}^{5}b^{7/2}},\quad L_{MK}=\frac{\bar{E}V_{0}}{2\bar{K}b^{3/2}}\label{tMK0}\\
\alpha & =1:\quad t_{MK}=\frac{4}{3}\frac{\bar{K}^{5}b^{7/2}}{\bar{\mu}\bar{E}^{4}Q_{0}^{2}},\quad L_{MK}=\frac{2}{3}\frac{\bar{K}^{4}b^{2}}{\bar{\mu}\bar{E}^{3}Q_{0}}\label{tMK1}
\end{align}

Corresponding solution conversion relations between the toughness
$k$ and transitional $MK$ scalings follow in the form of power laws
of the normalized time $t/t_{MK}$, 
\begin{equation}
\mathcal{M}=\mathcal{K}^{-5}=\left(\frac{t}{t_{MK}}\right)^{2\alpha-1},\qquad\frac{\gamma_{MK}}{\gamma_{k}}=\frac{L_{k}}{L_{MK}}=\left(\frac{t}{t_{MK}}\right)^{\alpha},\qquad\frac{\Omega_{MK}}{\Omega_{k}}=\frac{W_{k}}{W_{MK}}=1\label{Mvst}
\end{equation}

Transient solutions for power-law injection rescaled to the $MK$
scaling, and, therefore, showing explicit dependence of the crack
length and opening on time, are shown on Fig. \ref{fig:power'}. Particularly
for the opening, Fig. \ref{fig:power'}b clearly shows the increasing
/ diminishing opening with continuing crack propagation for `faster'
$\alpha>1/2$ and `slower' $\alpha<1/2$ injection power-laws $V\propto t^{\alpha}$,
respectively. 

At $\alpha=1/2$, the transitional scaling is singular, i.e. transition
time $t_{MK}$ and length $L_{MK}$ scales are clearly either zero
or infinite, see (\ref{scales_MK}). In this case, $\mathcal{M}$
is time-invariant constant, reflecting the stationarity (absence of
transition) of the fracture propagation regime and the self-similarity
of the corresponding solution. As the result, the solution in transitional
scaling for $\alpha=1/2$ lacks clear meaning, and the reader is referred
to the solution in a non-singular scaling, such as the toughness-scaling,
in Fig. \ref{fig:power}.

\subsubsection*{Accuracy of EoM solution}

Accuracy of the EofM approximate solutions can be gaged from Figs.
\ref{fig:power}-\ref{fig:power'}, which contrast the former, shown
by dashed lines, with the method-of-lines numerical solutions shown
by solid lines and characterized by the superior accuracy (sub 0.1\%
error, cf. Appendix \ref{App:error}). EofM accuracy is seen to depend
on the injection power-law $\alpha$. The relative error of approximation
of the crack half-length and opening generally does not exceed few
percent (and is particularly good for the constant injection rate
solution for which the error upper bound corresponds to that of the
zero-toughness solution limit discussed earlier, i.e. <1\% for length
and <5\% for the opening). More details on the EofM errors and their
estimation are given in Appendix \ref{App:error_EofM} and Fig. \ref{fig:power-error}. 

\subsection{Solution for Fluid Injection Followed by Shut-in\label{Sec:shut-in}}

In the shut-in problem, the fracture is first advanced by fluid injection
over a period of time, followed by the shut-in and subsequent propagation
with the fixed fracture volume. The fixed-volume fracture solution
of Section \ref{Sec:power} considers fracture propagation due to
an \emph{instantaneous} injection of a finite volume of fluid. In
the context of the shut-in problem, this asymptotic solution is therefore
expected to provide the large-time, post-shut-in asymptote, when the
details of the fracture evolution during the pre-shut-in period become
inconsequential. This asymptotic solution, as shown on Fig. \ref{fig:power'}
in red ($\alpha=0$), starts out of the viscosity-dominated regime
and ends up in the toughness-dominated regime. 

Full solution to the shut-in problem on a \emph{finite} shut-in timescale
considered here corresponds to an injection at \emph{constant rate}
which persists over finite duration $t<t_{0}$, followed by the shut-in
for $t\ge t_{0}$. Corresponding evolution of the dimensionless viscosity
is non-monotonic, increasing from zero (toughness-dominated regime)
at early time to the maximum value $\mathcal{M}_{0}=\mathcal{M}(t_{0})$
at the shut-in, and then decreasing asymptotically back to zero (fracture
arrest) at large time. Solution \emph{prior to the shut-in,} $t<t_{0}$,
has been given in Section \ref{Sec:power} using both simplified EoM
and the fully-numerical method-of-lines approaches and shown on Fig.
\ref{fig:power'} ($\alpha=1$). The \emph{post-shut-in} solution,
$t\ge t_{0}$, is obtained here using the same two approaches with
$t\dot{V}/V$ set to zero\footnote{see EofM (\ref{ODE'}) and equation (\ref{ODE_again}) in Appendix
\ref{App:num}, respectively.} and the `initial' conditions at the shut-in $t=t_{0}$ are those
of solution continuity, i.e. given by the aforementioned injection
solution up to that moment. 

\begin{figure}[tbph]
\begin{centering}
(a)\includegraphics[scale=0.45]{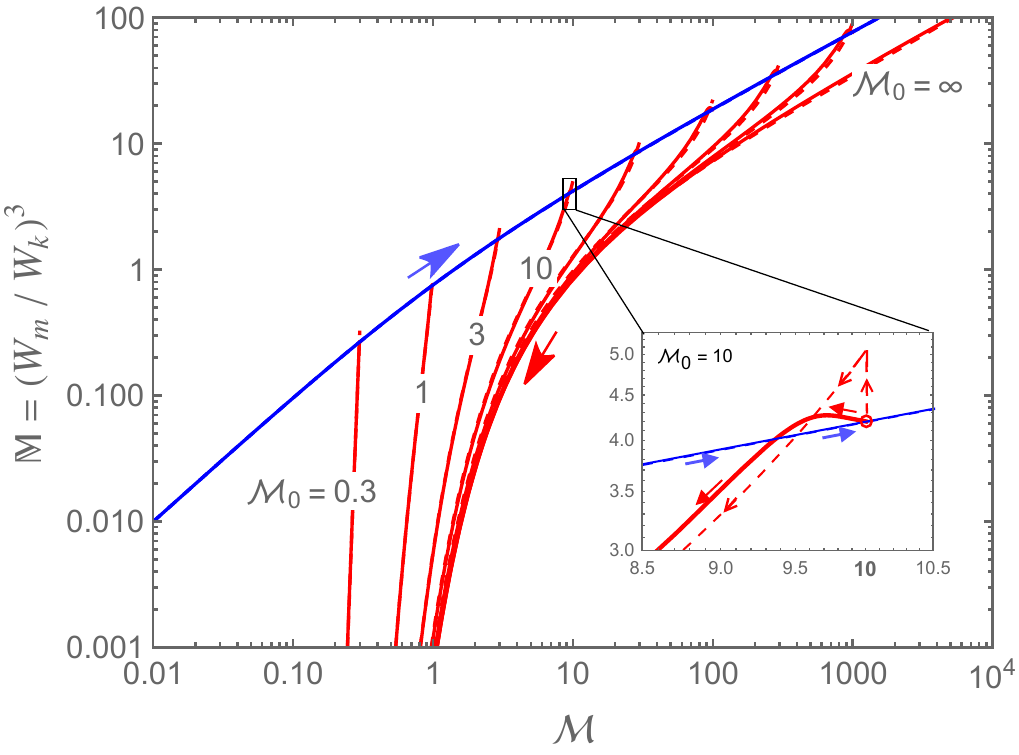}\medskip{}
\par\end{centering}
\begin{centering}
(b)\includegraphics[scale=0.45]{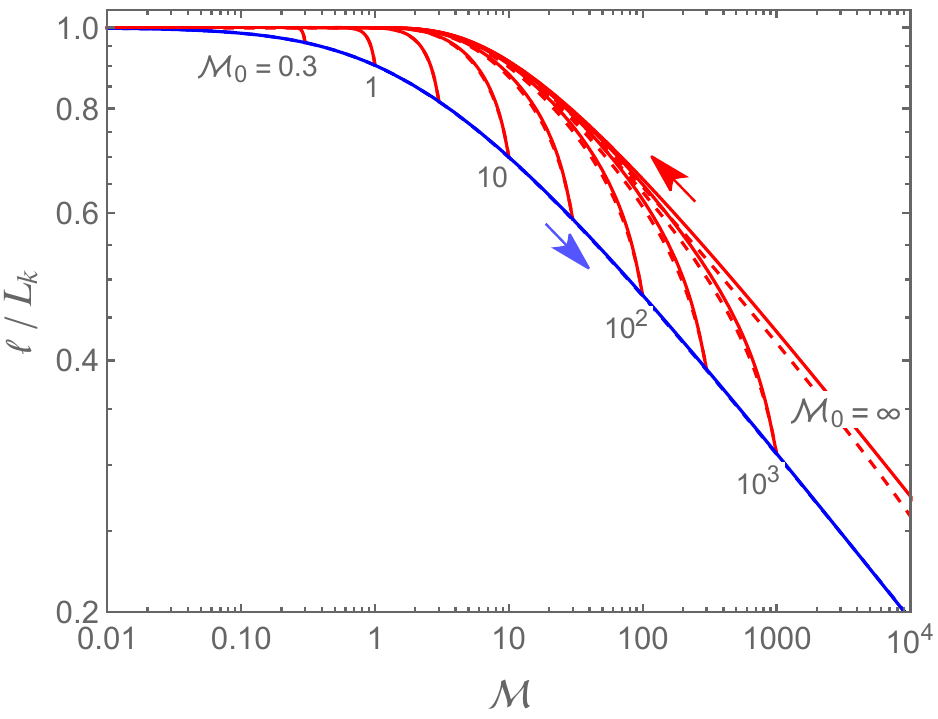}\quad{}(c)\includegraphics[scale=0.44]{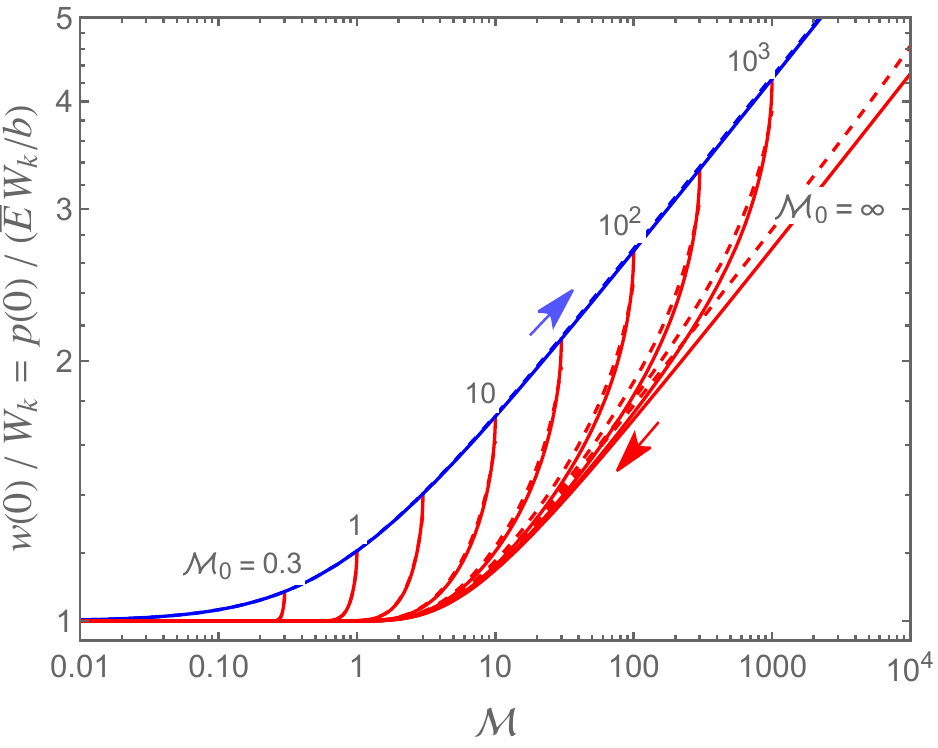}
\par\end{centering}
\caption{Shut-in problem. Equation-of-Motion (dashed) and the numerical Method-of-Lines
(solid) solutions for the hydraulic fracture propagation driven by
\emph{injection} at constant rate $V(t<t_{0})=Q_{0}t$ (blue) followed
by \emph{shut-in} $V(t\ge t_{0})=V_{0}=Q_{0}t_{0}$ (red) for various
values of the shut-in time $t_{0}$, as correspond to the values of
dimensionless viscosity parameter $\mathcal{M}_{0}=\mathcal{M}(t_{0})=(3\bar{\mu}\bar{E}^{4}/4\bar{K}^{5}b^{7/2})(V_{0}^{2}/t_{0})=$
0.3, 1, ... , 300, 1000. \textbf{(a)} Implicit (crack-propagation-dependent)
dimensionless viscosity parameter $\mathbb{M}\propto\ell\dot{\ell}$
evolution with the explicit dimensionless viscosity parameter $\mathcal{M}(t)\propto V^{2}(t)/t$.
\textbf{(b,c)} Corresponding evolution of the normalized crack half-length
$\ell/L_{k}(t)$ and opening/net-pressure at the inlet $w(0)/W_{k}=p(0)/(\bar{E}W_{k}/b)$
with $\mathcal{M}(t)$. Dependence of the solution on time follows
from that of the scales, $W_{k}=\bar{K}\sqrt{b}/\bar{E}$ and $L_{k}(t)=V(t)/(2bW_{k})$,
and of the evolution parameter $\mathcal{M}(t)=(3\bar{\mu}\bar{E}^{4}/4\bar{K}^{5}b^{7/2})(V^{2}(t)/t)$,
as further explored in Fig. \ref{fig:shut-in'}. The sense of time-evolution
of $\mathcal{M}(t)$ - increasing prior to the shut-in and decreasing
post the shut-in - is indicated by arrows. The injection solution
(blue) and the large-shut-in-time asymptote (red, $\mathcal{M}_{0}=\infty$)
correspond to the power-law $\alpha=1$ (constant-volume-rate) and
$\alpha=0$ (constant-volume) solutions in Fig. \ref{fig:power},
respectively. Inset in (a) zooms into the evolution of the solution
with $\mathcal{M}_{0}=10$ near the shut-in instant. \label{fig:shut-in}}
\end{figure}

Normalized solution for the implicit viscosity $\mathbb{M}\propto\ell\dot{\ell}$,
crack length, and inlet opening / net-pressure in the toughness scaling
is shown on Fig. \ref{fig:shut-in} for various values of the shut-in
time, as quantified by the corresponding values of dimensionless viscosity
parameter $\mathcal{M}_{0}=\mathcal{M}(t_{0})$. The injection part
of the solution, shown in blue, evolves, as previously discussed,
from the toughness- towards the viscosity-dominated regime with $\mathcal{M}(t)$
increasing with time, while the post-shut-in solution, shown in red,
signifies a reverse regime transition with $\mathcal{M}(t)$ now decreasing
with time, back towards toughness-dominated regime (asymptotic arrested
fracture state).
\begin{figure}[tbph]
\begin{centering}
(a)\includegraphics[scale=0.43]{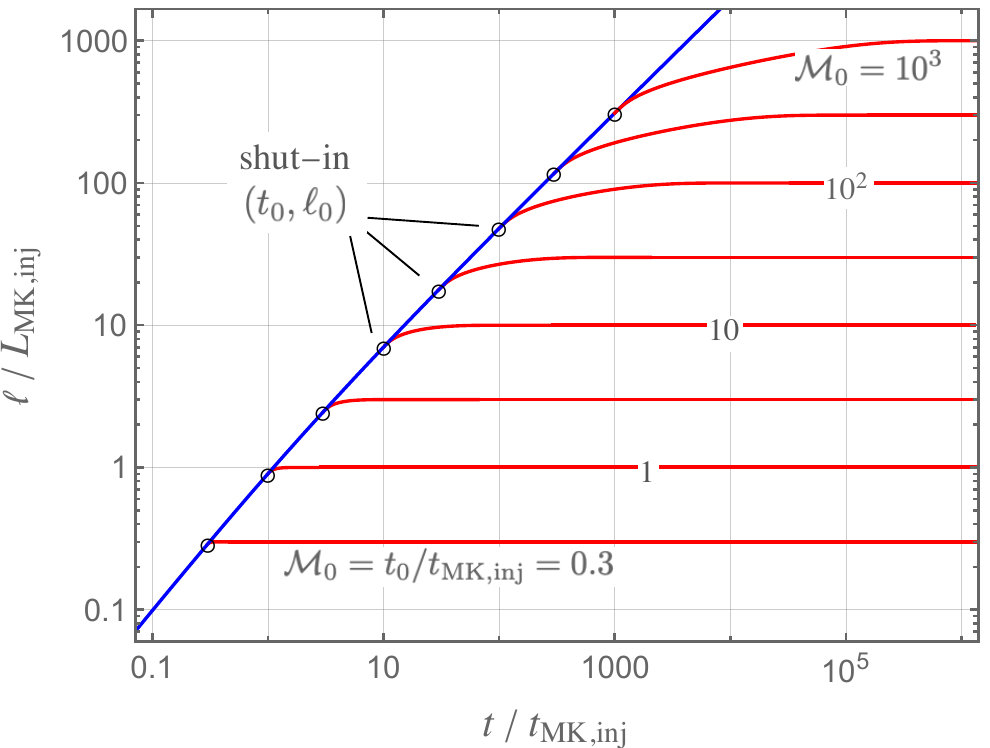}\quad{}(b)\includegraphics[scale=0.43]{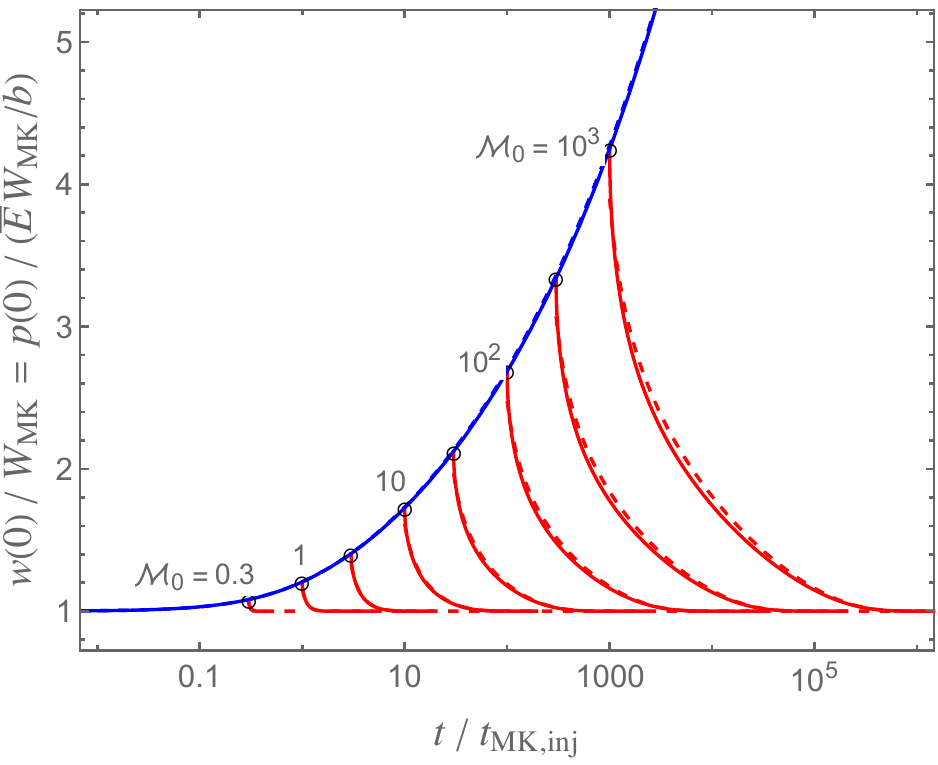}\vspace{0.5cm}
\par\end{centering}
\begin{centering}
(c)\includegraphics[scale=0.5]{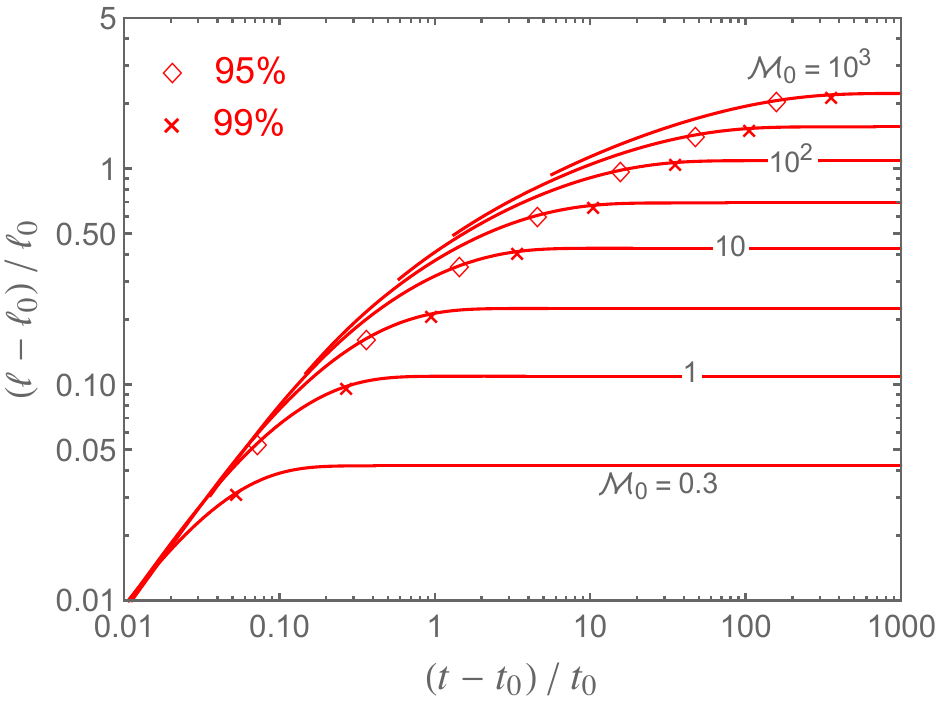}\quad{}(d)\includegraphics[scale=0.19]{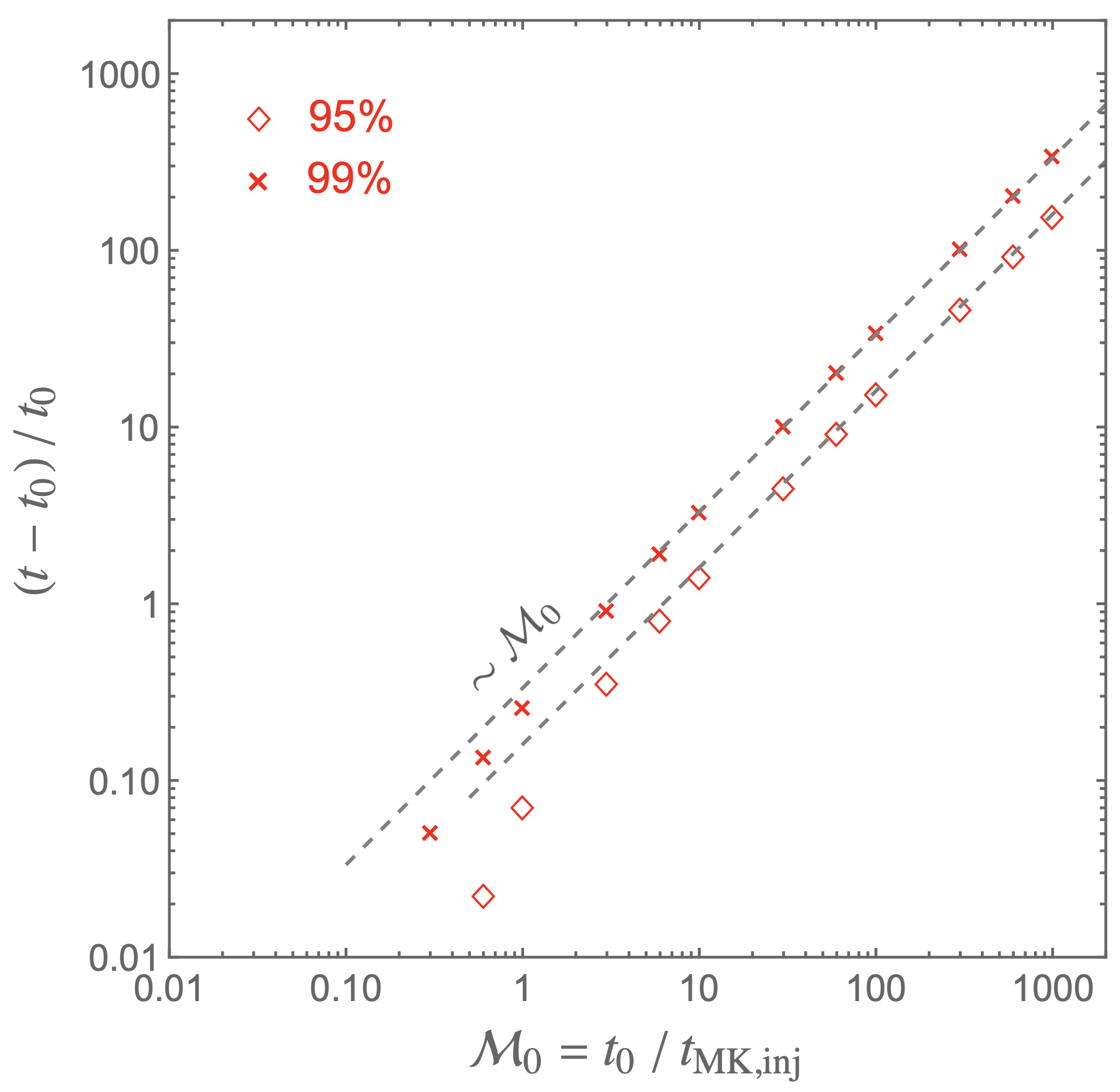}
\par\end{centering}
\caption{Shut-in problem. Solution for (a) crack half-length and (b) opening/net-pressure
of Fig. \ref{fig:shut-in}(b-c) re-scaled using the time-independent
scales, $\ell/L_{MK,\text{inj}}$ and $w(0)/W_{MK}$, and shown as
a function of scaled time $\tau=t/t_{MK,\text{inj}}$, where `inj'
indicates injection ($\alpha=1)$ transitional scale (i.e. $t_{MK,\text{inj}}=t_{MK}(\alpha=1)$
given by (\ref{tMK1})). (c) Post-shut-in crack growth, $\ell-\ell_{0}$
vs. $t-t_{0}$, normalized by the shut-in values, $\ell_{0}$ and
$t_{0}$. Symbols marks the states corresponding to $95\%$ and $99\%$
of the terminal fracture length $\ell_{max}=L_{k,V=V_{0}}=V_{0}/(2bW_{k})$.
(d) Normalized post-shut-in time $(t-t_{0})/t_{0}$ at $95\%$ and
$99\%$ of the terminal fracture length from (c) shown vs. the normalized
shut-in time $\mathcal{M}_{0}=t_{0}/t_{\text{MK,inj}}$. Dashed lines
show a linear approximation for large-enough shut-in time. \label{fig:shut-in'}}
\end{figure}

We note with regard to the EofM solution that the approximation for
non-dimensional crack length $\gamma_{k}\approx\varUpsilon(\mathbb{M})$
in the EofM framework is distinct for injection and post-shut-in (cf.
(\ref{omega})). Thus, continuity of EofM solution at $t=t_{0}$ can
only be enforced for \emph{either} fracture length $\gamma_{k}$ \emph{or}
dynamic non-dimensional parameter $\mathbb{M}$ (proportional to fracture
length and its time rate). EofM post-shut-in solutions presented here
use the fracture-length continuity and thus, expectedly, have a discontinuity
at $t=t_{0}$ in the fracture speed, as well as, in the crack opening,
the latter owing to the two distinct approximations for the opening
during injection and post-shut-in, respectively, in the EofM approach.
The inset in Fig. \ref{fig:shut-in}a zooms-in to show the discontinuity
of $\mathbb{M}\propto\ell\dot{\ell}$ in the EofM solution (dashed)
compared to the continuity of the method-of-lines solution.

Given the changing nature the time-dependence of $\mathcal{M}\propto V^{2}(t)/t$
and toughness lengthscale $L_{k}\propto V(t)$, and the different
nature of this time-dependence pre- and post- shut-in, we show the
shut-in solution rescaled to the time-independent, injection transitional
scaling in Fig. \ref{fig:shut-in'} (i.e. using scales $t_{\text{MK,inj}}$,
$L_{\text{MK,inj}}$, and $W_{\text{MK,inj}}=W_{k}$ where the index
`inj' indicates injection-scaling (\ref{tMK1})). Specifically, Fig.
\ref{fig:shut-in'}a-b show the evolution in time of the crack half-length
and inlet opening / net-pressure. We observe the post-shut-in continued
growth of the hydraulic fracture towards the terminal, toughness-dominated
state, with the extent of this growth an increasing function of the
dimensionless viscosity at the shut-in $\mathcal{M}_{0}$, which can
also be understood in terms of the increasing shut-in time $t_{0}$
and volume $V_{0}$. In other words, the further away is the fracture
propagation regime from being toughness-dominated at the shut-in of
the injection, the longer is the subsequent post-shut-in evolution
towards the fixed-volume toughness-dominated crack. This thesis is
further explored in Fig. \ref{fig:shut-in'}c showing the extent of
the post-shut-in crack growth (relative to the shut-in crack length)
vs. the post-shut-in duration (relative to the injection duration).
The symbols indicate the fracture within $1\%$ or $5\%$ of the terminal
length. Since the shut-in crack length $\ell_{0}$ is increasing with
the increased injection duration $t_{0}$ (see the injection solution
of Fig. \ref{fig:shut-in'}a shown in blue), the post-shut-in crack
growth $\ell-\ell_{0}$ is increasing faster than linearly with $\ell_{0}$,
Fig. \ref{fig:shut-in'}c. Corresponding normalized post-shut-in time
$(t-t_{0})/t_{0}$ for the fracture to reach within $1\%$ or $5\%$
of its final extent is shown as a function of the normalized shut-in
time $\mathcal{M}_{0}=t_{0}/t_{\text{MK,inj}}$ on Fig. \ref{fig:shut-in'}d.
Apart from the cases corresponding to small shut-in time (when the
pre-shut-in fracture propagates in/near the toughness-dominated regime,
$\mathcal{M}_{0}<1$), the Fig. suggests a linear scaling for the
normalized post-shut-in growth time $(t-t_{0})/t_{0}$ with $t_{0}$
shown by dotted lines. This translates to a quadratic dependence for
the dimensional post-shut-in time, e.g. the post-shut-in time to reach
95\% of the terminal length is $t_{95\%}-t_{0}\approx0.16\times t_{0}^{2}/t_{\text{MK,inj}}$
(Fig. \ref{fig:shut-in'}d). 

Finally, Fig. \ref{fig:shut-in'}b tracks the evolution of the inlet
opening / net-pressure with time, illustrating the inflation of the
crack at the inlet during the injection period, with subsequent deflation
post-shut-in towards the final values $W=W_{k}$ and $p=P_{k}=\bar{E}W_{k}/b$,
respectively. Since the fracture volume remains fixed post-shut-in,
the latter inlet deflation corresponds to the continued incremental
crack growth and redistribution of fluid within the crack, away from
the inlet. 

\section{Discussion and Conclusions}

The original PKN model of elongated hydraulic fracture propagation
\citep{PeKe61,Nord72} neglects the effect of rock toughness, in favor
of the dissipation in the viscous fluid flow inside the fracture.
In other words, energy dissipation in breaking the rock is assumed
negligible compared to that in the viscous fluid in the PKN elongated
fracturing. Energy considerations for hydraulic fractures of other
geometries (radial, plane-strain) have also seem to indicate that
the rock toughness, if constrained by the lab-scale measurements,
can be neglected in industrial hydraulic fracturing treatments, for
the representative values of the fracturing fluid viscosity, injection
rate and fracture scale \citep[e.g., ][]{Garagash09,Garagash11,Detournay16}.\textbf{ }

The extension of the PKN model to account for the rock fracture toughness
\citep{SarvaraminiGaragash15,chuprakov2017-PKN,dontsov22PKN} and
solutions thereof developed in this work allow to evaluate the effect
of rock toughness on different stages of HF propagation, during injection
and shut-in. In doing so, we can evaluate the effect of toughness
assuming either (i) typical values measured in the laboratory, small-scale
tests; or (ii) order(s) of magnitude larger, fracture-scale-dependent
values inferred, albeit not without controversy, from observations. 

\begin{table}[th]
\begin{centering}
{\footnotesize{}\caption{{\footnotesize{}An example of the regime-transition time $t_{MK}$
and length $L_{MK}$ scales for a typical slick-water hydraulic fracturing
treatment: during injection (regime transition $K\rightarrow M$)
and during the shut-in (transition towards the $K$-regime of the
terminal fracture) evaluated for three distinct realizations of the
fracture energy scaling with fracture height $G_{c}\propto b^{\chi}$,
Eq. (\ref{tough}): $\chi=0$ (scale-invariant), $0.5$ (sub-linear),
$1$ (linear). The values of the non-dimensional viscosity (regime
parameter) $\mathcal{M}_{0}=\mathcal{M}(t_{0})$ and fracture half-length
$\ell(t_{0})$ at the end of injection are also given. Parameter set
for a field elongated fracture (Fig. \ref{fig1}b,c) was used: fracture
height $2b\approx0.1$ km, injection rate, $Q\approx100$ L/s ($t<t_{0}\approx2.8$
hr, $2V(t)<2V_{o}\approx10^{3}$ m$^{3}$) and $Q=0$ ($t\ge t_{0}$),
fluid viscosity of fracturing gel $\mu=150$ cP, and the values of
the rock parameters, $E'=30$ GPa and $K_{Ic,0}=1$ MPa$\sqrt{\text{m}},$
corresponding to small-scale ($2b_{0}\sim0.1$m) laboratory measurements
on a reservoir sandstone \citep[e.g. ][]{chandler2016fracture}.}
\label{tab:param}}
}{\footnotesize\par}
\par\end{centering}
\begin{centering}
\par\end{centering}
\centering{}%
\begin{tabular}{|c|c|c|c|c|c|c|c|}
\hline 
\multirow{1}{*}{{\footnotesize{}Frac. energy scaling}} & {\footnotesize{}Frac. toughness} & \multicolumn{2}{c|}{{\footnotesize{}Injection ($K\rightarrow M$)}} & \multicolumn{4}{c|}{{\footnotesize{}Shut-in ($\,\rightarrow K$)}}\tabularnewline
\cline{3-8} \cline{4-8} \cline{5-8} \cline{6-8} \cline{7-8} \cline{8-8} 
{\footnotesize{}$G_{c}\propto b^{\chi}$} & {\footnotesize{}MPa$\sqrt{\text{m}}$} & {\footnotesize{}$t_{\text{MK}}$, hr} & {\footnotesize{}$L_{\text{MK}}$, km} & {\footnotesize{}$\mathcal{M}(t_{0})$} & {\footnotesize{}$\ell(t_{0})$, km} & {\footnotesize{}$t_{\text{MK}}$, hr} & {\footnotesize{}$L_{\text{MK}}\!=\!\ell(\infty)$, km}\tabularnewline
\hline 
\hline 
{\footnotesize{}$\chi=0$ (scale-invariant)} & {\footnotesize{}$1$} & {\footnotesize{}$\approx0$} & {\footnotesize{}$\approx0$} & {\footnotesize{}$\sim10^{6}$} & {\footnotesize{}$0.6$} & {\footnotesize{}$\sim10^{6}$} & {\footnotesize{}$8.5$}\tabularnewline
\hline 
{\footnotesize{}$\chi=0.5$ (sub-linear)} & {\footnotesize{}$5.6$} & {\footnotesize{}$0.005$} & {\footnotesize{}$0.005$} & \textbf{\footnotesize{}$300$} & {\footnotesize{}$0.6$} & \textbf{\footnotesize{}$400$} & {\footnotesize{}$1.5$}\tabularnewline
\hline 
{\footnotesize{}$\chi=1$ (linear)} & {\footnotesize{}$32$} & {\footnotesize{}$27$} & {\footnotesize{}$5$} & \textbf{\footnotesize{}$0.05$} & {\footnotesize{}$0.27$} & {\footnotesize{}$0.07$} & {\footnotesize{}$0.27$}\tabularnewline
\hline 
\end{tabular}
\end{table}

Effects of toughness and its scaling with the fracture height on the
dominant propagation regime of a hydraulic fracture are illustrated
in Table \ref{tab:param} for the values of problem parameters representative
of a gel (high-viscosity) fracturing treatment in Cartage Cotton Valley
gas reservoir \citep{Rutledge04,Mayerhofer00}.. The dominant regimes
can be ascertained by comparing the regime-transition time (and length)
scales to the injection time $t_{0}\approx2.8$ hr and fracture length
$\ell(t_{0})$, respectively. 

Assuming scale-invariance ($\chi=0$) of the fracture toughness (i.e.
the field fracture toughness is given by the small-scale laboratory
value $K_{Ic}=K_{Ic,0}=1$ MPa$\sqrt{\text{m}}$), we observe that
the injection stage is viscosity-dominated, i.e. the time $t_{\text{MK}}$
to transition from the `early-time' toughness ($K$) to the `large-time'
viscosity ($M$) dominated regime is vanishingly small. Corresponding
non-dimensionless viscosity evaluated at the end of injection is exceedingly
large, underlining irrelevance of rock toughness during the injection
stage for the considered case. Consequently, the post shut-in propagation
from the at-shut-in $M$ to the eventual $K$ regime would occur over
exceedingly large, impractical timescale $\sim10^{6}$ hr, with the
corresponding forecasted post-shut-in fracture growth from $0.6$
km to $8.5$ km for the half-length. The fracturing fluid leak-off
is expected to become dominant over large post-shut-in time \citep{dontsov22PKN},
which would effectively curtail the growth predicted here by the zero-leak-off
model. However, even assuming the leak-off mediated final fracture
half-length approximately equal to that at the shut-in, the theoretical
prediction for the latter $0.6$ km is unrealistic when contrasted
to the final fracture half-length value $\ell_{\text{final}}\approx0.4$
km inferred from observations (Fig. \ref{fig1}b,c).

A very different picture of the fracture evolution emerges for a scale-dependent
fracture toughness. Considering the linear fracture energy scaling
with height ($\chi=1$), the corresponding field fracture toughness
in this example, $K_{Ic}\approx32$ MPa$\sqrt{\text{m}}$, is more
than an order of magnitude larger than the laboratory value, which
leads to the toughness-dominated injection and post-shut-in propagation.
Indeed, the time $t_{\text{MK}}$ to transition from the `early-time'
toughness ($K$) to the `large-time' viscosity ($M$) dominated regime
during \emph{injection} is exceedingly large in this case (much larger
than the injection time period). This is further highlighted by very
small value of the non-dimensional viscosity at the end of injection,
$\mathcal{M}(t_{0})\approx0.05$, and negligible post-shut-in propagation
over a very short post-shut-in transition timescale ($0.07$ hr).
The final predicted fracture half-length $0.27$ km is comparable
to the observed $0.4$ km (Fig. \ref{fig1}b,c). 

Consequently, we surmise based on this typical field example that
hydraulic fracture propagation is viscosity-dominated, and therefore
adequately approximated by the zero-toughness solution (i.e. the original
PKN model \citep{PeKe61,Nord72}), when a typical laboratory (small-scale)
fracture toughness value $\sim1$ MPa$\sqrt{\text{m}}$ is assumed.
The opposite is true, i.e. the hydraulic fracture propagation is predicted
to be toughness-dominated, and therefore adequately approximated by
the zero-viscosity solution, if scale-dependence of the fracture toughness
is invoked. This conclusion arrived at in the example of high-viscosity
gel fluid fracturing will certainly hold for low-viscosity (e.g. `slick'
water) fracturing.

Examination of a larger well-constrained dataset for field hydraulic
fracture and dikes and model inversion thereof is needed to further
validate the scale dependence argument. 

\pagebreak{}

\bibliographystyle{elsarticle-harv}
\bibliography{bibrefs_garagash_Aug21}

\pagebreak{}

\appendix

\section{Numerical Method-of-Lines Solution\label{App:num}}

\setcounter{equation}{0}
\numberwithin{equation}{section}
\numberwithin{figure}{section}

In this Appendix we provide a method-of-line numerical solution method
for a PKN fracture propagation. We solve the problem in the \emph{toughness
scaling}, specifically for the normalized opening $\Omega$ as a function
of normalized coordinate $\xi$ and time $t$ (or alternatively dimensionless
viscosity $\mathcal{M}(t)$). We approximate the cube of the opening
$\Omega^{3}(t,\xi)$ in space by piecewise linear distribution\footnote{Linear approximation for $\Omega^{3}$ is asymptotically correct at
the tip} defined by the values of the opening $\Omega_{i}(t)=\Omega(t,\xi_{i})$
at the uniformly spaced set of $n+1$ nodes $\xi_{i}=i/n$ with $i=0,...n$. 

Boundary conditions for the crack opening and fluid velocity at the
crack tip dictate, respectively,
\begin{equation}
\Omega_{n}=1,\quad\mathbb{M}=-\frac{\partial\Omega^{3}}{\partial\xi}_{\left|\xi=1\right.}\approx-\frac{3\Omega_{n}^{3}-4\Omega_{n-1}^{3}+\Omega_{n-2}^{3}}{2/n}\label{tip''}
\end{equation}

Carrying out the crack volume integral using the aforementioned piecewise
linear approximation for $\Omega^{3}$, the global continuity yields
\begin{equation}
\frac{1}{\gamma}\approx\frac{3}{4/n}\sum_{i=0}^{n-1}\frac{\Omega_{i+1}^{4}-\Omega_{i}^{4}}{\Omega_{i+1}^{3}-\Omega_{i}^{3}}\label{global''}
\end{equation}
Alternatively, an inlet boundary condition can be used in place of
the global continuity, as follows from (\ref{inlet'}), substituting
$t\dot{\ell}/\ell=\mathcal{\mathbb{M}}/(\mathcal{M}\gamma^{2})$,
and simplifying
\begin{equation}
\frac{t\dot{V}}{V}\mathcal{M}\gamma=-\frac{3}{4}\frac{\partial\Omega^{4}}{\partial\xi}_{\left|\xi=0\right.}\approx\frac{3}{4}\frac{3\Omega_{0}^{4}-4\Omega_{1}^{4}+\Omega_{2}^{4}}{2/n}\label{inlet''}
\end{equation}

Combining local continuity (\ref{cont'}) with the Poiseuille law,
1st in (\ref{Ktip}), and substituting $t\dot{\ell}/\ell=\mathcal{\mathbb{M}}/(\mathcal{M}\gamma^{2})$
yields upon rearranging the double-derivative term:
\[
t\dot{\Omega}-\frac{\mathcal{\mathbb{M}}}{\mathcal{M}\,\gamma^{2}}\left(\xi\frac{\partial\Omega}{\partial\xi}+\frac{3}{4\mathcal{\mathbb{M}}}\frac{\partial^{2}\Omega^{4}}{\partial\xi^{2}}\right)=0
\]
Which approximation over the internal $n-1$ nodes using 2nd order
central finite differences for spatial derivatives yields $n-1$ ODEs
\begin{equation}
\frac{d\Omega_{i}}{d\ln t}-\frac{\mathcal{\mathbb{M}}}{\mathcal{M}\,\gamma^{2}}\left(\xi_{i}\frac{\Omega_{i+1}-\Omega_{i-1}}{2/n}+\frac{3}{4\mathcal{\mathbb{M}}}\frac{\Omega_{i+1}^{4}-2\Omega_{i}^{4}+\Omega_{i-1}^{4}}{1/n^{2}}\right)=0\quad(i=1,...,n-1)\label{lines}
\end{equation}
The closing ODE is rendered by plugging $t\dot{\ell}/\ell=t\dot{V}/V+t\dot{\gamma}/\gamma$
into relation $t\dot{\ell}/\ell=\mathcal{\mathbb{M}}/(\mathcal{M}\gamma^{2})$
\begin{equation}
\frac{t\dot{V}}{V}+\frac{d\ln\gamma}{d\ln t}=\frac{\mathcal{\mathbb{M}}}{\mathcal{M}\,\gamma^{2}}\label{ODE_again}
\end{equation}

Equations (\ref{lines}) and (\ref{ODE_again}) provide a system of
$n$ ODEs to solve for the evolution in time of $n$ unknowns $\Omega_{0},...,\Omega_{n-1}$
in lieu of the known $\Omega_{n}=1$ and the above expressions for
$\mathbb{M}$, 2nd in (\ref{tip''}), and $\gamma$, either (\ref{global''})
or (\ref{inlet''}).

Once again we can choose to use dimensionless viscosity $\mathcal{M}$
as a 'time' by exchanging logarithmic time derivative in above for
(\ref{ODE'}). As discussed in the main text, this is particularly
convenient when injected fluid volume is a given power law of time,
i.e. $t\dot{V}/V=\alpha=const$. 

The initial conditions for an injection problem with $\alpha>1/2$
correspond to the toughness-dominated conditions and are prescribed
at some small initial value of $\mathcal{M}=\mathcal{M}_{\text{ini}}\ll1$
by the small-viscosity solution (\ref{small-M}). In the case when
$\alpha<1/2$, the initial conditions corresponds to the viscosity-dominated
conditions, and zero-toughness solution, expressed in the toughness
scaling, (\ref{large-M}), at some large $\mathcal{M}=\mathcal{M}_{\text{ini}}\gg1$
is used to initialize the solution.

The shut-in part of the crack evolution for $t\ge t_{0}$ ($\mathcal{M}\le\mathcal{M}_{0}$)
is solved for using the same framework with $t\dot{V}/V$ set to zero
and 'initial' conditions at $t=t_{0}$ given by the injection solution
up to that moment.

Numerical solutions presented in the main text and figures use the
global continuity approximation for $\gamma$, (\ref{global''}),
and are carried using $n=50$ spatial elements. Numerical error and
convergence are discussed in Appendix \ref{App:error}.

\section{Numerical Error Estimates}

\begin{figure}[tbph]
\begin{centering}
\includegraphics[scale=0.5]{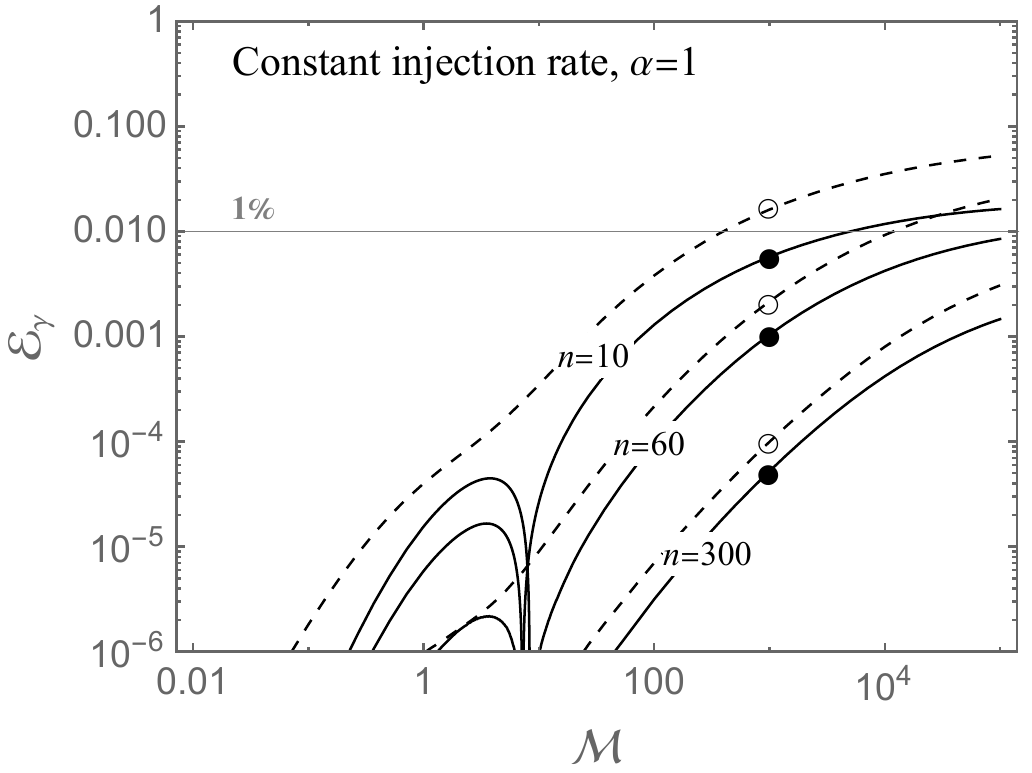}\includegraphics[scale=0.49]{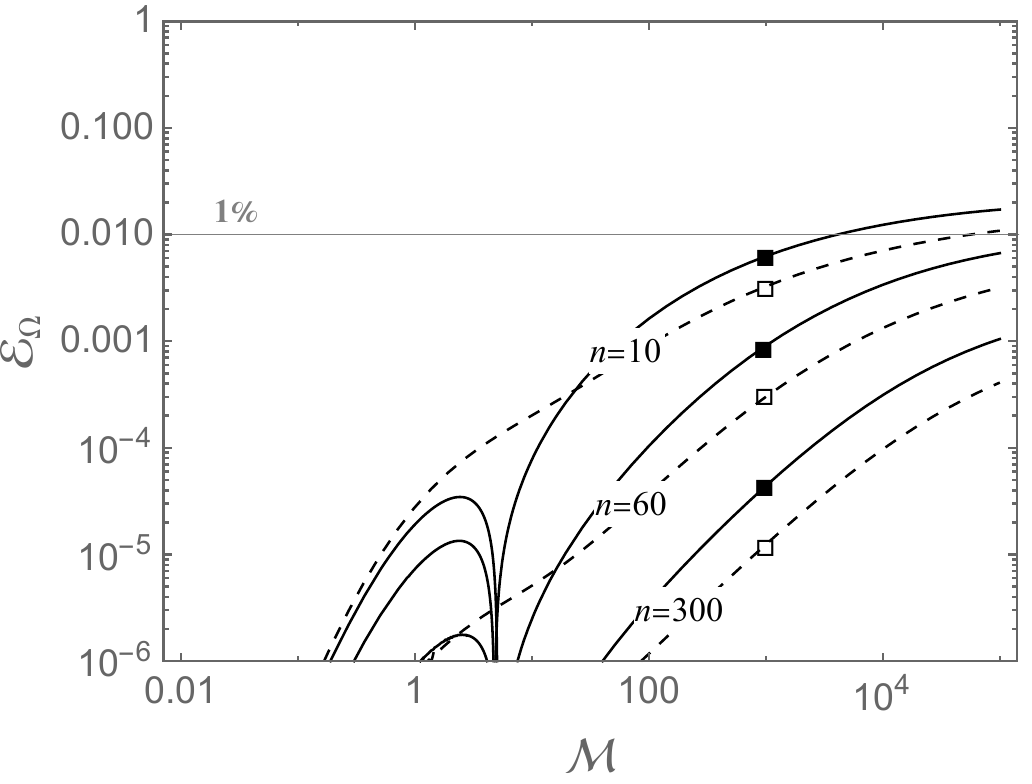}
\par\end{centering}
\caption{Method-of-lines error for solutions with various discretization $n$
in the \emph{constant-injection-rate} case, $\alpha=t\dot{V}/V=1$,
as a function of the dimensionless time $\mathcal{M}=t/t_{MK}(\alpha=1)$.
\textbf{(a)} crack length error relative to $n=600$ solution, $\mathcal{E}_{\gamma}=|\gamma(\mathcal{M})_{n}/\gamma(\mathcal{M})_{600}-1|$,
and \textbf{(b)} crack opening at the inlet error $\mathcal{E}_{\Omega}=|\Omega(\mathcal{M},0)_{n}/\Omega(\mathcal{M},0)_{600}-1|$.
Solid and dashed lines correspond to the method-of-lines implementations
using global (\ref{global''}) and local, at the inlet, (\ref{inlet''})
continuity conditions, respectively. \label{fig:num-conv}}
\end{figure}
\begin{figure}[tbph]
\begin{centering}
\includegraphics[scale=0.5]{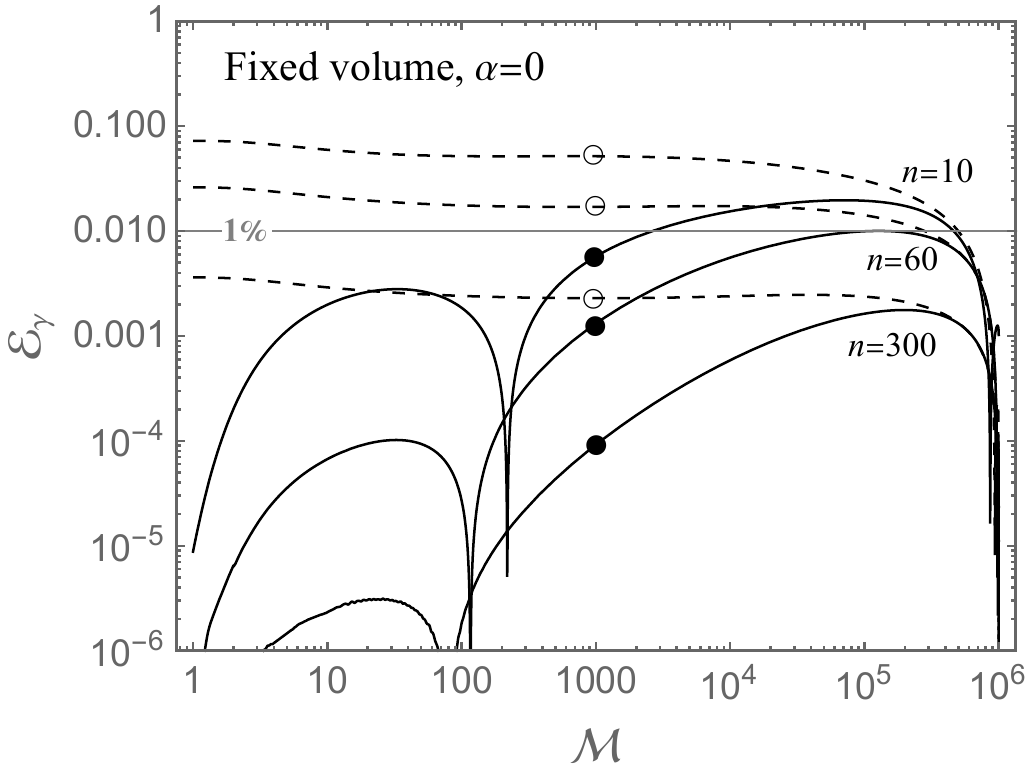}\includegraphics[scale=0.49]{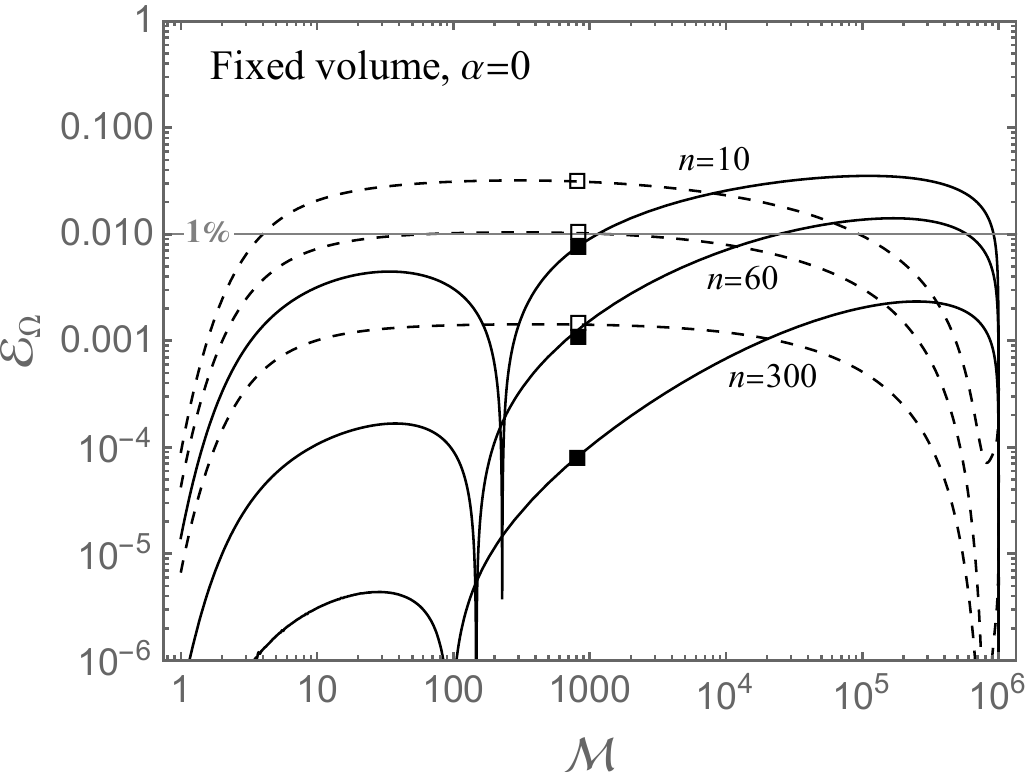}
\par\end{centering}
\caption{Same as Fig. \ref{fig:num-conv} but for the \emph{fixed-volume} injection
case $\alpha=0$. The evolution parameter $\mathcal{M}$ in this case
is a dimensionless reciprocal of time, $\mathcal{M}=t_{MK}(\alpha=0)/t$.
\label{fig:num-conv-0}}
\end{figure}
\begin{figure}[tbph]
\begin{centering}
\includegraphics[scale=0.5]{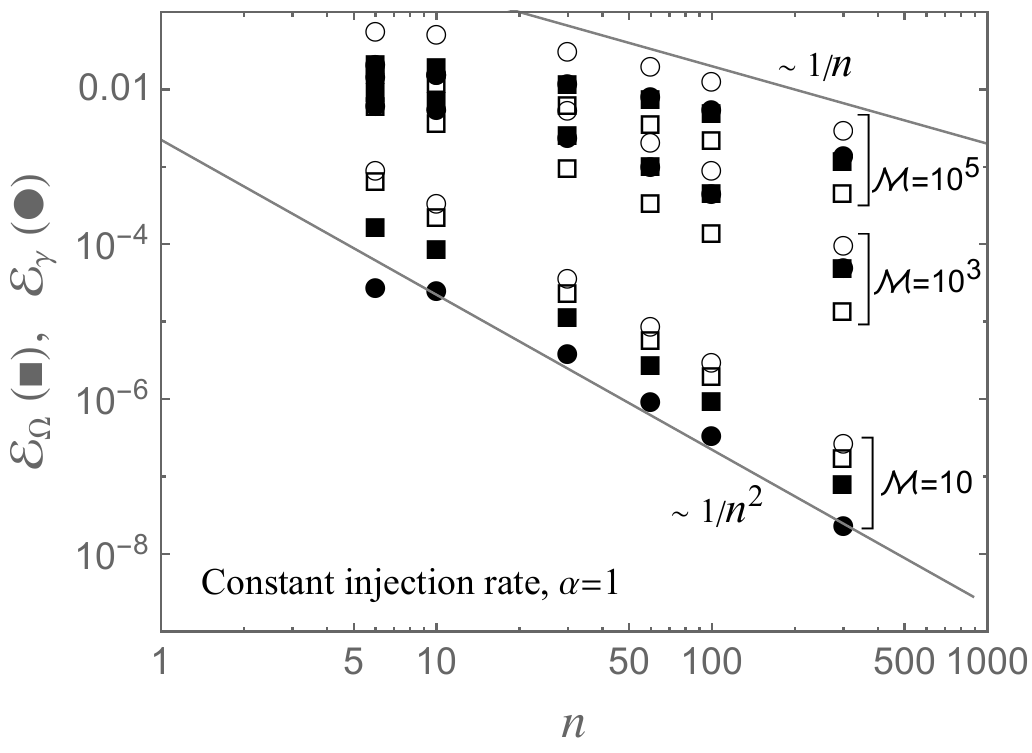}\includegraphics[scale=0.5]{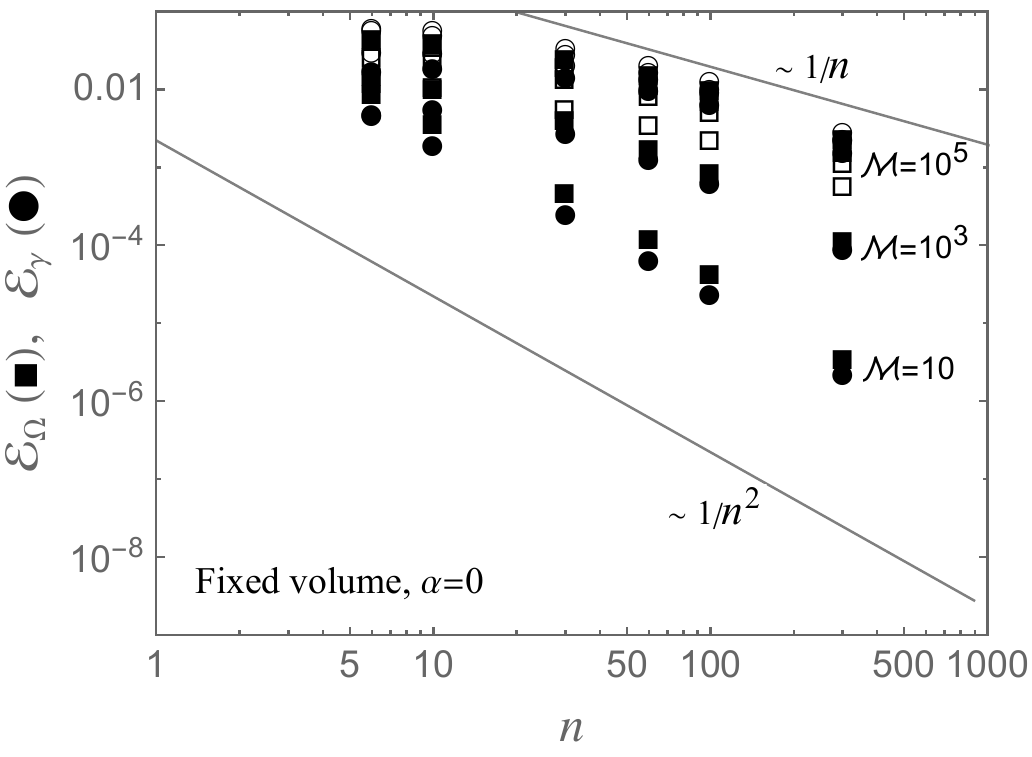}
\par\end{centering}
\caption{Convergence of the method-of-lines solution with increasing spatial
discretization $n$ at fixed values of the evolution parameter $\mathcal{M}=10,10^{3},10^{5}$
in the case of \textbf{(a)} constant injection rate, $\alpha=1$,
and \textbf{(b)} fixed volume injection, $\alpha=0$. Solid and empty
symbols correspond to the method-of-lines implementations using global
(\ref{global''}) and inlet (\ref{inlet''}) continuity conditions,
respectively. The method utilizing the global continuity, in both
constant-rate and fixed-volume cases, converges as $\sim1/n^{2}$
at intermediate values of $\mathcal{M}=10,10^{3}$, and slower, as
$\sim1/n$ in the viscosity-dominated regime $\mathcal{M}=10^{5}$.
On the other hand, the method utilizing the local-inlet continuity
has slower $\sim1/n$ in the fixed-volume case, (b), in the entire
range of $\mathcal{M}$. \label{fig:num-conv-1}}
\end{figure}

\subsection{Method-of-Lines\label{App:error}}

The integration of the ODE system (\ref{lines}) and (\ref{ODE_again})
in the method-of-lines numerical solution (Appendix \ref{App:num})
is carried using adaptive-stepping routine in Wolfram Mathematica,
and is numerically `exact' for all practical purposes (i.e. near the
machine precision). Method's numerical error is therefore associated
solely with the spatial discretization and can be estimated by considering
the solution convergence with increasing number $n$ of the spatial
nodes, Figs. \ref{fig:num-conv}-\ref{fig:num-conv-1}. Figs. \ref{fig:num-conv}
and \ref{fig:num-conv-0} show, for two injection scenarios, evolution
in `time' $\mathcal{M}$ of the errors in crack length and opening
value at the inlet for solutions at various discretization $n$. in
two cases of constant-rate-injection, $\alpha=1$, and fixed-volume,
$\alpha=0$, respectively. The error is defined as the relative distance
of the solution for a given $n$ from the one with the largest value
of $n$ considered ($n=600$). The error of the method with the global
continuity implementation (solid lines and solid symbols in the Figures)
is seen to increase with $\mathcal{M}$ towards the viscosity-dominated
limit for both constant-rate and fixed-volume injection cases, where
this limit is reached at large and small times, respectively. This
remains true for the method with the local-inlet continuity implementation
(dashed lines and open symbols) for the constant-rate case (Fig. \ref{fig:num-conv}),
but the error remains approximately insensitive to the value of $\mathcal{M}$
in the fixed-volume case (\ref{fig:num-conv-0}). (Abrupt decrease
of the error near the largest value of $\mathcal{M}=10^{6}$ in the
fixed-volume solutions is an artifact of the initialization of the
solution in this case). Overall, when both injection cases considered,
the method based on global continuity is seen to result in lower error
compared to the method based on the local-inlet continuity. Fig. \ref{fig:num-conv-1},
which shows solutions' errors, sampled at few values of $\mathcal{M}$,
vs. spatial discretization $n$, illustrates the convergence of the
numerical methods.

\subsection{Equation-of-Motion\label{App:error_EofM}}

\begin{figure}[tbh]
\begin{centering}
\includegraphics[scale=0.5]{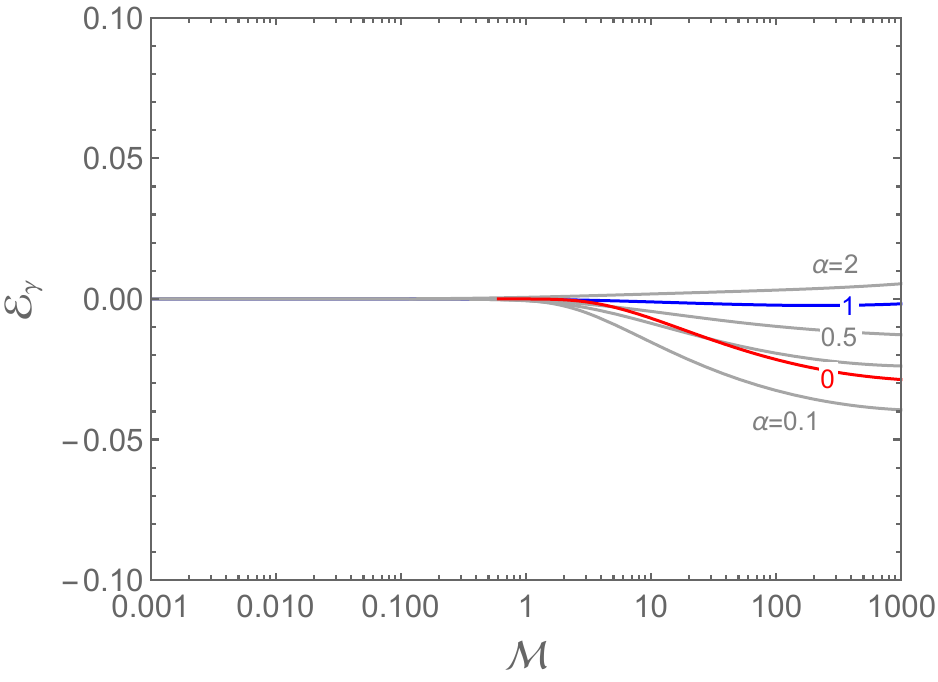}\includegraphics[scale=0.5]{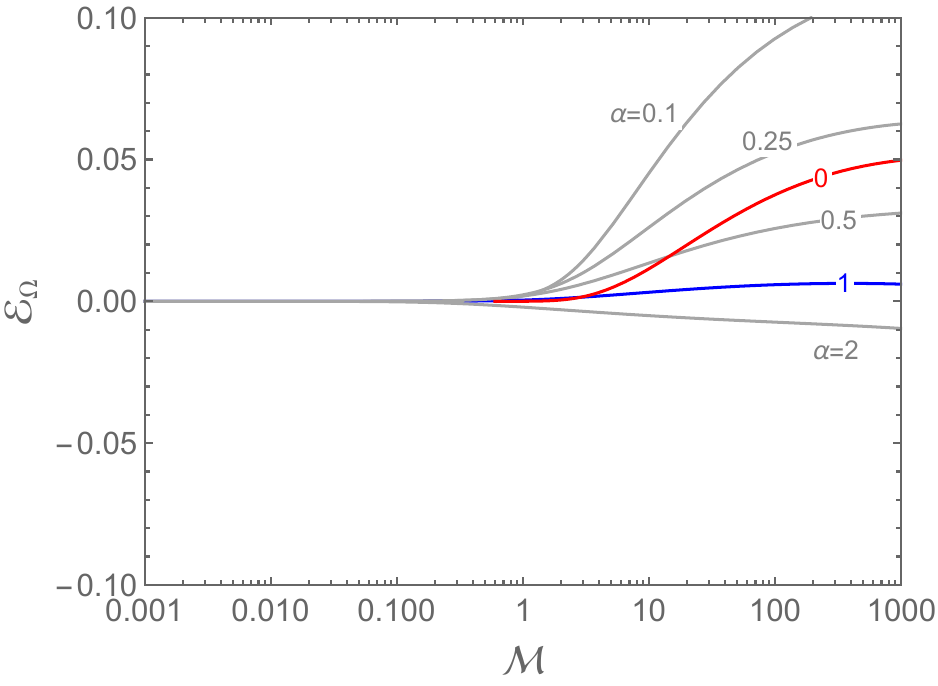}
\par\end{centering}
\caption{Relative error of the EofM solutions, $\mathcal{E}_{\gamma}=\gamma_{\text{EofM}}/\gamma_{\text{MofL}}-1$
and $\mathcal{E}_{\Omega}=\Omega_{\text{EofM}}(0)/\Omega_{\text{MofL}}(0)-1$,
for various injection power-laws $\alpha=t\dot{V}/V$ as function
of evolutionary variable $\mathcal{M}=(t/t_{MK}(\alpha))^{2\alpha-1}$.
Both the EofM and the full numerical, method-of-lines (MofL) solutions
with $n=50$ are as shown in Fig. \ref{fig:power}. \label{fig:power-error}}
\end{figure}
Approximation error of an EofM solution is considered relative to
the corresponding fully numerical, method-of-lines solution with $n=50$,
and illustrated as a function of dimensionless evolution variable
$\mathcal{M}$ on Fig. \ref{fig:power-error} for various injection
power-laws $\alpha$.  Method-of-lines solutions are indeed very
accurate (error of solutions with $n=50$ is bounded by $\sim0.001$
($0.1\%$) in the considered range of $\mathcal{M}$, see Figs. \ref{fig:num-conv}-\ref{fig:num-conv-1}),
and, thus, appropriate to serve as a reference in defining the EofM
solution error. It is observed, e.g., that the EofM solution for the
case of constant injection rate is very accurate with the error (shown
in linear scale) discernible only as solution approaches the viscosity-dominated
regime, where it remains sub $1\%$. The error is seen to increase
significantly in the viscosity-affected part of the solution for small
injection power-law values $\alpha$ (see for example the error for
the lowest-non-zero value $\alpha=0.1$ shown). Error of the EofM
solution for the fixed-volume case, $\alpha=0$, is bounded by $3\%$
for the crack length and $5\%$ for the opening at the inlet. Possibly
surprising result that error for $\alpha=0$ case is less than for
the cases with $\alpha=0.1$ and $\alpha=0.25$ is understood upon
recalling different opening spatiotemporal approximation that has
been used in the EofM construction for injection $\alpha>0$ (\ref{inj_approx})
and the fixed-volume $\alpha=0$ (\ref{shut-in_approx}) cases. 
With the former approximation deteriorating with decreasing power-law
$\alpha$.

\end{document}